\documentclass[a4paper,11pt]{article}
\usepackage{jcappub} 
\usepackage{orcidlink}
\usepackage{makecell}
\usepackage{amsmath}
\usepackage{longtable}
\usepackage{multirow}
\usepackage{romannum}
\usepackage{booktabs}  
\usepackage{threeparttable}  
\usepackage{adjustbox}

\usepackage{tikz}
\usepackage{placeins}
\usetikzlibrary{arrows.meta,positioning,calc}

\title{\boldmath Lyman-$\alpha$ forest constraints on\\
pure and mixed fuzzy dark matter}

\author[a, b]{Jianxiang Liu~\orcidlink{0009-0004-4092-3931},}
\author[a,b,c,1]{Yan Gong~\orcidlink{0000-0003-0709-0101},\note{Corresponding author.}}
\author[a]{and Xingchen Zhou~\orcidlink{0000-0001-7283-1100}}

\affiliation[a]{National Astronomical Observatories, Chinese Academy of Sciences, Beijing, 100101, China}
\affiliation[b]{School of Astronomy and Space Sciences, University of Chinese Academy of Sciences, Beijing, 100049, China}
\affiliation[c]{Science Center for Chinese Space Station Survey Telescope, National Astronomical Observatories, Chinese Academy of Sciences, Beijing 100101, China}

\emailAdd{liujx@bao.ac.cn} 
\emailAdd{gongyan@bao.ac.cn}
\emailAdd{xczhou@nao.cas.cn}

\abstract{Fuzzy dark matter (FDM), often realized as an ultralight scalar field, can suppress the growth of small-scale structures and could be strictly tested with Lyman-$\alpha$ forest measurements.
In this work, we constrain both pure and mixed FDM models (PFDM and MFDM) using measurements of the one-dimensional (1D) Lyman-$\alpha$ forest flux power spectrum at $z=5.0$, 4.6, and 4.2.
We perform cosmological hydrodynamical simulations with modified initial conditions and construct a two-stage neural network emulator for accurate analysis.
The first stage predicts the cold dark matter (CDM) 1D flux power spectrum, while the second stage predicts the MFDM effect relative to the CDM baseline.
This construction improves the sensitivity to weak FDM effects, enforces the correct CDM limit, and enables robust interpolation across a broad range of FDM masses and fractions.
After marginalizing over the intergalactic medium parameters, we obtain the FDM mass $m_{\mathrm{FDM}}>1.9\times10^{-21}~\mathrm{eV}$ at 95\% credible level for the PFDM model.
For the MFDM model, we find the FDM fraction of dark matter $f_{\mathrm{FDM}}<0.07$, $0.12$, and $0.65$ at 95\% credible level for $\log_{10}(m_{\mathrm{FDM}}/\mathrm{eV})=-23.0$, $-22.0$, and $-21.0$, respectively.
When $\log_{10}(m_{\mathrm{FDM}}/\mathrm{eV})\gtrsim -20$, the current data do not provide an effective upper limit on $f_{\mathrm{FDM}}$.
}

\begin{document}
\maketitle
\flushbottom

\section{Introduction} \label{sec:1}
Dark matter (DM) is a hypothetical form of matter that constitutes most of the matter content in the Universe~\cite{2020A&A...641A...6P}. 
Although its existence is supported by multiple probes, such as the cosmic microwave background (CMB~\cite{2020A&A...641A...1P}) and galaxy rotation curves~\cite{1970ApJ...159..379R,1985ApJ...295..305V,2015PASJ...67...75S}, its fundamental nature remains unknown, making it one of the most important open questions in modern cosmology.
The cold dark matter (CDM) model, exemplified by weakly interacting massive particles (WIMPs), has been highly successful in explaining the large-scale structure of the Universe~\cite{1996PhR...267..195J,2005Natur.435..629S,2012AnP...524..507F,2014MNRAS.444.1518V,2018MNRAS.475..624N}. However, the lack of a definitive experimental detection~\cite{2025JCAP...11..080T,2025arXiv251212176W}, together with several challenges on small scales~\cite{2017Galax...5...17D,2017ARA&A..55..343B,2026NatAs..10..440V, 2026arXiv260116818H}, has motivated the proposal and reconsideration of a wide range of non-cold dark matter models~\cite{2000PhRvL..85.1158H,2016PhR...643....1M,2021ARA&A..59..247H,2025arXiv250700705E,2019PrPNP.104....1B,2001PhLB..518....8B, 2018PhRvL.121h1301G,2018PhR...730....1T,2025RvMP...97d5004A}.

Among the non-cold dark matter models, fuzzy dark matter (FDM) is one of the most attractive candidates~\cite{2000PhRvL..85.1158H,2016PhR...643....1M,2021ARA&A..59..247H,2025arXiv250700705E}.
In its canonical form, FDM consists of ultralight bosons with mass $m_\mathrm{FDM} \sim 10^{-22}~\mathrm{eV}$.
Such a small mass gives rise to wave-like behavior and an effective quantum pressure, which can suppress the growth of structure and the matter power spectrum on small scales~\cite{2000PhRvL..85.1158H,2016ApJ...818...89S,2022PhRvD.105l3529P,2023MNRAS.524.4256M}.
Although the pure FDM model (PFDM), where all dark matter is described by a single ultralight scalar field, is now under increasing pressure from issues such as the so-called Catch-22 problem, {namely the tension that smaller $m_\mathrm{FDM}$ produces larger and less dense cores while simultaneously more strongly suppressing small-scale matter clustering}~\cite{2020MNRAS.492.5721D,2019PhRvL.123e1103M,2021PhRvL.126g1302R,2025PhRvL.134o1001Z,2025ApJ...986..127N,2026ApJ..1000...88L}, no definitive conclusion has been reached{~\cite{2022PhRvD.106f3517D,2023MNRAS.521.2608M,2025arXiv250902781M,2025MNRAS.540.2653C,2026arXiv260307175W,2026arXiv260426393L,2026PhRvD.113b3055T,2026arXiv260401278C}}.
A natural and important extension of PFDM is the case of multiple fields~\cite{2016PhR...643....1M,2023PhRvD.107h3014G,2023MNRAS.524L..84P}.
A particularly relevant scenario is mixed FDM (MFDM), in which only a fraction of dark matter is composed of a light FDM component, while the remaining component behaves effectively as CDM on the scales of interest.
This scenario can significantly alleviate the tension between PFDM and current observations{~\cite{2023JCAP...06..023R,2024ApJ...976...40W,2024PhRvD.110l3532L,2026arXiv260610006C,2026arXiv260606410L,2026JCAP...01..047V}}.

{To explore the properties of dark matter with high precision, it is essential to accurately characterize the matter distribution, especially on small scales.}
The Lyman-$\alpha$ forest consists of absorption lines of neutral hydrogen in the spectra of high-redshift quasars~\cite{2016ARA&A..54..313M,2026enap....4..401T}.
These absorption features arise from fluctuations in the intergalactic medium (IGM) around mean cosmic density.
Since the IGM traces the underlying matter distribution, the one-dimensional (1D) flux power spectrum of the Lyman-$\alpha$ forest encodes information about matter clustering over a wide range of scales.
This makes the Lyman-$\alpha$ forest a powerful probe of dark matter models.
Using this observable for dark matter constraints generally requires hydrodynamical simulations, since the mapping between the matter density field and the transmitted flux is non-linear and depends on the thermal and ionization state of the IGM~\cite{2013PhRvD..88d3502V,2017PhLB..773..258G,2017PhRvD..96b3522I,2017MNRAS.471.4606A,2017PhRvD..96l3514K,2017PhRvL.119c1302I,2018PhRvD..98h3540M,2021PhRvL.126g1302R,2022JCAP...10..032H,2023PhRvD.108b3502V,2024PhRvD.109d3511I,2025PhRvD.112d3502G,2026arXiv260324331Z,2026arXiv260304401G}.
On large and mildly non-linear scales, effective field theory provides a useful complementary approach, while fully non-linear small-scale analyses still rely on simulations~\cite{2024PhRvD.109b3507I,2026PhRvL.136q1402I}.

In this work, we present a simulation-based analysis of PFDM and MFDM using the 1D flux power spectrum of the high-redshift Lyman-$\alpha$ forest~\cite{2019ApJ...872..101B}.
The data span the redshift range $z = 4.2$--$5.0$ and are measured from high-resolution quasar spectra obtained with VLT/UVES~\cite{2000SPIE.4008..534D} and Keck/HIRES~\cite{1994SPIE.2198..362V}.
We first consider PFDM as a baseline case, since it has been extensively studied and provides a useful reference for comparison with previous work and for understanding parameter degeneracies.
We then extend the analysis to MFDM, which is a more general and physically well-motivated scenario.

To perform parameter inference in the multidimensional space of dark matter and IGM parameters, we run a suite of cosmological hydrodynamical simulations and construct a two-stage neural network emulator for the 1D flux power spectrum.
The first stage predicts the CDM baseline 1D flux power spectrum, while the second stage emulates the relative effect of PFDM or MFDM on top of this baseline.
By isolating the dark matter signal from the baseline, this two-stage structure provides highly accurate and computationally efficient predictions across the complex joint space of dark matter and IGM parameters.
Using this emulator in a statistical inference framework, we derive constraints on both PFDM and MFDM.

The paper is organized as follows: in section~\ref{sec:2}, we describe the observational data and the physical models of the IGM and FDM; the hydrodynamical simulations and the construction of the emulator are presented in section~\ref{sec:3}; in section~\ref{sec:4}, we show the constraints on PFDM and MFDM; in section~\ref{sec:5}, we summarize our results and discuss future prospects.
Throughout this work, we adopt the best-fitting $\it Planck$ 2018 cosmological parameters $h=0.6732$, $\Omega_\mathrm{m}h^2=0.14314$ and $\Omega_\mathrm{b}h^2=0.022383$~\cite{2020A&A...641A...6P}.

\section{Data and physical models} \label{sec:2}

\subsection{Observational data}\label{sec:2.1}
We use the 1D flux power spectrum from ref.~\cite{2019ApJ...872..101B} at redshifts $z=5.0$, 4.6, and 4.2.
The measurements are estimated from 15 high-resolution quasar spectra observed with VLT/UVES~\cite{2000SPIE.4008..534D} and Keck/HIRES~\cite{1994SPIE.2198..362V}.
At each redshift, the 1D flux power spectrum is measured as a function of the line-of-sight velocity wavenumber $k_\mathrm{f}$.
The data cover the range
$-2.2 \leq \log_{10}\left(k_\mathrm{f}/\mathrm{s}\,\mathrm{km}^{-1}\right) \leq -0.7$,
with 16 data points at each redshift.
This gives 48 data points in total, with typical relative uncertainties of about 10\%--25\%.

{Following the procedure described in ref.~\cite{2024PhRvD.109d3511I}, we use the resolution-corrected flux power spectrum provided by ref.~\cite{2019ApJ...872..101B}. We then apply the corresponding resolution correction to the uncorrected covariance matrix provided by ref.~\cite{2019ApJ...872..101B}.}
The highest wavenumber used in this work reaches $k_\mathrm{f} \simeq 0.2~\mathrm{s}\,\mathrm{km}^{-1}$, which roughly corresponds to a comoving wavenumber of $k > 20~h\,\mathrm{Mpc}^{-1}$.
The extension to these small scales makes the data particularly sensitive to the suppression of structure formation, and therefore can provide strong constraints on non-cold dark matter models.

\subsection{IGM thermal and ionization model}\label{sec:2.2}
To estimate the 1D flux power spectrum from the Lyman-$\alpha$ forest~\cite{1997ApJ...486..599H}, we first define the flux contrast in the line-of-sight velocity space as
\begin{equation}
\label{eq:deltaf}
\delta_\mathrm{f}(v) = \frac{F(v)}{\langle F \rangle} - 1,
\end{equation}
where $F(v)=\exp[-\tau(v)]$ is the transmitted flux at velocity $v$, $\tau(v)$ is the optical depth at the same velocity, and $\langle F \rangle$ is the mean transmitted flux at a given redshift.
We then decompose the transmitted-flux field along each sightline into Fourier modes, $\tilde{\delta}_\mathrm{f}(k_\mathrm{f})$. 
The 1D flux power spectrum is defined by the variance of these Fourier modes, and we have
\begin{equation}
\label{eq:fluxpower}
P_\mathrm{f}(k_\mathrm{f}) \propto \left\langle |\tilde{\delta}_\mathrm{f}(k_\mathrm{f})|^2 \right\rangle ,
\end{equation}
where the average is taken over all sightlines.

The 1D flux power spectrum, $P_\mathrm{f}(k_\mathrm{f})$, is sensitive to the thermal state of the IGM, and 
several thermal processes can affect $P_\mathrm{f}(k_\mathrm{f})$ on small scales.
First, Doppler broadening caused by gas thermal velocities suppresses the 1D flux power spectrum at large $k_\mathrm{f}$.
A higher gas temperature leads to broader absorption features and therefore stronger suppression of small-scale power.
Most of the IGM gas around mean cosmic density can be described by a power-law temperature--density relation~\cite{1997MNRAS.292...27H},
\begin{equation}
\label{eq:Tandd}
T(z,\Delta) = T_0(z)\Delta^{{\gamma}(z)-1},
\end{equation}
where $\Delta = \rho/\bar{\rho}$ is the gas overdensity, with $\rho$ and $\bar{\rho}$ denoting the gas density and the mean gas density, respectively.
This temperature--density relation has two free parameters: the temperature at mean density, $T_0(z)$, and the slope, ${\gamma}(z)$. 
These parameters provide a useful description of the thermal broadening effect on the 1D flux power spectrum.

Pressure smoothing and gas peculiar velocities also have impacts on the small-scale 1D flux power spectrum~\cite{1998MNRAS.296...44G,2002MNRAS.334..107G,2003ApJ...583..525G,2015ApJ...812...30K}.
Unlike thermal broadening, pressure smoothing depends on the integrated thermal history of the IGM rather than on the instantaneous gas temperature~\cite{2016MNRAS.463.2335N}.
Gas peculiar velocities also can be affected by the dynamical response of the gas to this thermal history, although their detailed dependence is more complex~\cite{2024PhRvD.109d3511I}.
A useful parameter related to these effects is the integrated energy injected per unit mass at the mean density, $u_0(z)$, which is defined as~\cite{2007MNRAS.374..493B,2016MNRAS.463.2335N} 
\begin{equation}
\label{eq:u0}
u_0(z) =
\int_{z_\mathrm{end}}^{z_\mathrm{ini}}
\frac{\mathrm{d}z'}{H(z')(1+z')}
\frac{\sum_i n_i \epsilon_i}{\bar{\rho}},
\end{equation}
where $n_i$ and $\epsilon_i$ are the number density and photoheating rate of species $i \in [\mathrm{H\,I}, \mathrm{He\,I}, \mathrm{He\,II}]$, and $H(z)$ is the Hubble parameter. 
The integration range $[z_\mathrm{end},z_\mathrm{ini}]$ is chosen such that $u_0$ is most strongly correlated with the 1D flux power spectrum. 
{We use $[z_{\rm end}, z_{\rm ini}]=[6,13]$ at $z=5.0$, $[4.6,13]$ at $z=4.6$, and $[4.2,12]$ at $z=4.2$, as listed in Table~4 of ref.~\cite{2019ApJ...872..101B}.}

In addition, the mean transmitted flux, $\langle F \rangle$, is related to the hydrogen photoionization rate, $\Gamma_{\mathrm{HI}}$. 
Since the effective optical depth is defined as $\tau_\mathrm{eff} = -\ln \langle F \rangle$, it approximately satisfies $\tau_\mathrm{eff} \propto \Gamma_{\mathrm{HI}}^{-1}$ after reionization. 
We therefore marginalize over the uncertainty in the ionization state of the IGM by rescaling the effective optical depth~\cite{2005MNRAS.357.1178B}.
This rescaling is a computationally inexpensive but effective post-processing step. 
In this work, we adopt the fiducial redshift evolution~\cite{2019ApJ...872..101B}
\begin{equation}
\label{eq:taueff}
\tau_\mathrm{eff}^\mathrm{fid}(z)
=
1.56
\left(
\frac{1+z}{5.75}
\right)^4 ,
\end{equation}
and allow an overall rescaling parameter, $\tau_0(z)$, around this relation. In this way, the effective optical depth can be rescaled as $\tau_\mathrm{eff}^\mathrm{fid}(z)\times \tau_0(z)$.
Together, $T_0(z)$, $\gamma(z)$, $u_0(z)$, and $\tau_0(z)$ provide a compact description of the thermal and ionization state of the IGM relevant to the 1D flux power spectrum.

\subsection{Pure and mixed FDM models}\label{sec:2.3}
As discussed above, PFDM and MFDM can suppress the growth of structure and the matter power spectrum on small scales. For MFDM, its properties can be described by two parameters: the FDM particle mass, $m_\mathrm{FDM}$, and the FDM fraction, $f_\mathrm{FDM}$.
PFDM is a special case of MFDM with $f_\mathrm{FDM}=1$, while the CDM limit is recovered when $f_\mathrm{FDM}=0$.
To quantify the effect of MFDM on the linear matter power spectrum relative to CDM, we introduce the transfer function for the linear matter power spectrum, $\mathcal{T}(k)$, which is defined as~\cite{2001ApJ...556...93B}
\begin{equation}
\label{eq:matter_transfer}
\mathcal{T}^2(k)
=
\frac{
P_\mathrm{MFDM}(k)
}{
P_\mathrm{CDM}(k)
},
\end{equation}
where $P_\mathrm{MFDM}(k)$ and $P_\mathrm{CDM}(k)$ are the linear matter power spectra in the MFDM and CDM models, respectively.
The mass $m_\mathrm{FDM}$ mainly determines the scale at which the suppression begins, while the fraction $f_\mathrm{FDM}$ mainly determines the asymptotic plateau height of the transfer function at small scales~\cite{2017PhRvD..96l3514K}.
These two parameters together with the cosmological parameters determine the detailed shape of $\mathcal{T}(k)$.

Since the 1D flux power spectrum of the Lyman-$\alpha$ forest encodes information about matter clustering over a wide range of scales, we also introduce the transfer function for the 1D flux power spectrum, $\mathcal{T}_\mathrm{f}(k_\mathrm{f})$, to quantify the effect of MFDM relative to CDM.
It is defined as~\cite{2018PhRvD..98h3540M,2022JCAP...10..032H}
\begin{equation}
\label{eq:flux_transfer}
\mathcal{T}_\mathrm{f}^2(k_\mathrm{f})
=
\frac{
P_{\mathrm{f,MFDM}}(k_\mathrm{f})
}{
P_{\mathrm{f,CDM}}(k_\mathrm{f})
},
\end{equation}
where $P_{\mathrm{f,MFDM}}(k_\mathrm{f})$ and $P_{\mathrm{f,CDM}}(k_\mathrm{f})$ are the 1D flux power spectra in the MFDM and CDM models, respectively. 
It is important to note that $\mathcal{T}_\mathrm{f}(k_\mathrm{f})$ is not determined only by the dark matter and cosmological parameters.
The mapping from the matter distribution to the transmitted flux is sensitive to the thermal and ionization state of the IGM.
As a result, $\mathcal{T}_\mathrm{f}(k_\mathrm{f})$ also depends on IGM parameters such as $T_0(z)$, $\gamma(z)$, $u_0(z)$, and $\tau_0(z)$.

In practice, interpreting the small-scale 1D flux power spectrum in MFDM models requires hydrodynamical simulations.
In these simulations, the difference between MFDM and CDM is introduced through the modified initial conditions described by Eq.~\eqref{eq:matter_transfer}, namely by modifying the initial matter power spectrum~\cite{2017MNRAS.471.4606A,2017PhRvD..96l3514K,2017PhRvL.119c1302I,2021PhRvL.126g1302R}.
However, during the subsequent evolution, the difference between MFDM and CDM is not fully captured by the modified initial conditions, since the quantum pressure of the FDM component can also introduce additional suppression~\cite{2000PhRvL..85.1158H,2016ApJ...818...89S,2022PhRvD.105l3529P,2023MNRAS.524.4256M}, which we do not explicitly include in the hydrodynamical simulations for computational feasibility.

Note that this approximation is sufficiently accurate for PFDM over the parameter range relevant to current Lyman-$\alpha$ forest constraints~\cite{2019PhRvD..99f3509L,2019MNRAS.482.3227N}.
For MFDM with small $m_\mathrm{FDM}$, our constraints are expected to be conservative in this case.
This is because the CDM-like component clusters more efficiently into dense structures, so the local FDM fraction around the mean density IGM can be higher than the global fraction.
As a result, quantum pressure may still produce a noticeable additional suppression even when the global value of $f_\mathrm{FDM}$ is small~\cite{2026arXiv260406038W}.
Since our simulations attribute the MFDM-induced suppression entirely to the modified initial conditions, including the additional suppression from quantum pressure would tend to shrink the allowed MFDM parameter region.
In particular, the upper limits on $f_\mathrm{FDM}$ at fixed $m_\mathrm{FDM}$ would become stronger.
{Overall, for MFDM, modelling its effect only through modified initial conditions, without including the full Schr\"odinger--Poisson evolution, is a non-trivial assumption.
This assumption is expected to make our constraints conservative, but the precise magnitude of this effect remains uncertain and needs further investigation.}

\section{Simulations and emulator} \label{sec:3}
When using hydrodynamical simulations to interpret the small-scale 1D flux power spectrum in PFDM or MFDM models, parameter inference in these models involves sampling a high-dimensional parameter space that includes both dark matter and IGM parameters.
It is therefore computationally infeasible to run a hydrodynamical simulation for every parameter combination explored in the inference.
A widely used and efficient approach is to train an emulator on a finite set of simulations that cover the relevant parameter space~\cite{2013PhRvD..88d3502V,2021PhRvD.103d3526R,2025PhRvD.112d3502G}.
In sections~\ref{sec:3.1} and~\ref{sec:3.2}, we describe the hydrodynamical simulations and the construction of the emulator, respectively.

\subsection{Hydrodynamical simulations}\label{sec:3.1}
To study the effects of different dark matter and IGM parameters on the 1D flux power spectrum, we perform cosmological hydrodynamical simulations of the Lyman-$\alpha$ forest using the publicly available code \texttt{MP-Gadget}\footnote{\url{https://github.com/MP-Gadget/MP-Gadget/tree/master}}~\cite{yu_feng_2018_1451799,2022MNRAS.512.3703B,2022MNRAS.513..670N}.
We adopt a box size of $10~\mathrm{Mpc}\,h^{-1}$ with $2\times512^3$ dark matter and gas particles, corresponding to dark matter and gas particle masses of $5.48\times10^5~M_\odot\,h^{-1}$ and $1.02\times10^5~M_\odot\,h^{-1}$, respectively.
{In previous analyses based on the data of ref.~\cite{2019ApJ...872..101B}, this choice of box size and particle mass has been widely used~\cite{2021PhRvL.126g1302R,2024PhRvD.109d3511I,2025PhRvD.112d3502G,2021PhRvD.103d3526R,2026arXiv260628482Y}.
We test numerical convergence by varying the box size at fixed particle mass and by varying the particle mass at fixed box size.
We find that the 1D flux power spectrum is converged to within $10\%$ over most of the relevant range, with deviations not exceeding $15\%$ even in the most extreme cases.
We have further verified that this level of convergence is sufficient for obtaining stable constraints.}
All simulations start at $z=99$ from initial conditions generated with \texttt{MP-GenIC}\footnote{\url{https://github.com/MP-Gadget/MP-Gadget/tree/master/genic}}~\cite{2020JCAP...06..002B}, where dark matter and gas particles are initialized on offset grids.
The simulations are evolved to $z=4.2$, with snapshots saved at $z=5.0$, $4.6$, and $4.2$.

We include the effect of MFDM through the modified initial conditions, as described by the transfer function of the linear matter power spectrum in Eq.~\eqref{eq:matter_transfer}.
The corresponding linear matter power spectra are computed using the modified Boltzmann code \texttt{axionCAMB}\footnote{\url{https://github.com/dgrin1/axioncamb}}~\cite{2000ApJ...538..473L,2015PhRvD..91j3512H}.
For computational efficiency, we use the \texttt{Quick\_Lya} option, which converts gas particles with overdensity $\Delta>1000$ and temperature $T<10^5~\mathrm{K}$ into collisionless particles~\cite{2004MNRAS.354..684V}.

In the simulations, we adopt an optically thin and spatially uniform UV background~\cite{2012ApJ...746..125H}.
To obtain different thermal states of the IGM, we follow the reionization model of ref.~\cite{2017ApJ...837..106O}.
We vary the mid-point redshift of hydrogen reionization, $z_{\mathrm{rei}}$, to generate different default UV backgrounds.
We then modify the simulation input parameters $H_{\mathrm{A}}$ and $H_{\mathrm{S}}$ to introduce an overdensity-dependent rescaling of the default heating rates:
$\epsilon_i(z) = H_{\mathrm{A}}\,\epsilon_{0,i}(z)\,\Delta^{H_{\mathrm{S}}}$, for $i \in [\mathrm{H\,I}, \mathrm{He\,I}, \mathrm{He\,II}]$.
We further rescale the optical depth to match an effective optical depth of $\tau_\mathrm{eff}^\mathrm{fid}\times \tau_0$ (see section~\ref{sec:2.2}), thereby producing different ionization states of the IGM ~\cite{2021PhRvL.126g1302R,2024PhRvD.109d3511I,2025PhRvD.112d3502G}.
To compute the 1D flux power spectrum from the snapshots, we use \texttt{fake\_spectra}\footnote{\url{https://github.com/sbird/fake_spectra}}~\cite{2017ascl.soft10012B} to generate 10000 mock spectra and calculate the corresponding 1D flux power spectrum.

In addition, we apply corrections for patchy reionization and $\mathrm{Si\,III}$ following the methods of refs.~\cite{2022MNRAS.509.6119M,2026MNRAS.546f2262M}.
For the patchy reionization correction, we directly adopt the coefficients listed in Table~2 of ref.~\cite{2022MNRAS.509.6119M} without introducing an explicit dependence on the redshift of reionization.
This choice is motivated by two considerations.
First, our hydrodynamical simulations assume photoionization equilibrium and do not solve the full non-equilibrium equations used in ref.~\cite{2022MNRAS.509.6119M}.
Photoionization equilibrium is commonly adopted in Lyman-$\alpha$ forest analyses and provides sufficient accuracy for the results considered in this work~\cite{2019MNRAS.490.1588G,2021PhRvL.126g1302R}.
However, the detailed mapping between the thermal history and the physical reionization history can differ from that obtained in full non-equilibrium treatments.
Therefore, the reionization parameter in our simulations should not be interpreted as a direct counterpart of the reionization history parameter in these non-equilibrium models.
Second, the variation of the correction coefficients with the reionization redshift is smaller than the estimated $\sim$ 5\% uncertainty of the correction itself.
We have also verified that including or excluding these corrections has little impact on our final constraints.

As discussed above, our simulations are described by five parameters.
Two of them specify the MFDM model: the particle mass, $\log_{10}(m_\mathrm{FDM}/\mathrm{eV})\in[-23,-19]$, and the FDM fraction, $f_\mathrm{FDM}\in[0,1]$.
The other three parameters describe the input thermal history of the IGM: $H_{\mathrm{A}}\in[0.05,4]$, $H_{\mathrm{S}}\in[-1,1]$, and $z_\mathrm{rei}\in[6,15]$.
By varying these parameters, we obtain simulation outputs at each snapshot that span different dark matter models and IGM thermal states.
In the post-processing stage, we also vary the optical depth rescaling parameter $\tau_0$ from $0.3$ to $1.8$ in steps of $0.05$.
This gives 31 1D flux power spectra for each snapshot.
To cover the parameter space as broadly as possible with a limited number of simulations, we use Latin hypercube sampling to generate 100 training points and 10 independent test points in this five-dimensional parameter space~\cite{mckay2000comparison}.
{For the 100 training points, the mean nearest-neighbor distance in the normalized 5D hypercube is 0.288.
This indicates that the sampling is relatively sparse.
However, as we show below, these points are sufficient to construct an effective emulator.
In addition, the 10 independent test points include the cases that are most relevant for our analysis, namely small $m_{\mathrm{FDM}}$ with small $f_{\mathrm{FDM}}$, and large $m_{\mathrm{FDM}}$ with large $f_{\mathrm{FDM}}$.}
For each of the 100 training points, we run a pair of simulations.
One simulation uses the full five-parameter MFDM model, while the other is the corresponding CDM model with the same input thermal history parameters with $f_\mathrm{FDM}=0$.
These paired simulations are used to isolate the MFDM-induced relative change in the 1D flux power spectrum.
For the 10 independent test points, we only run the full five-parameter MFDM simulations.

\begin{figure}
    \centering
    \includegraphics[width=\linewidth]{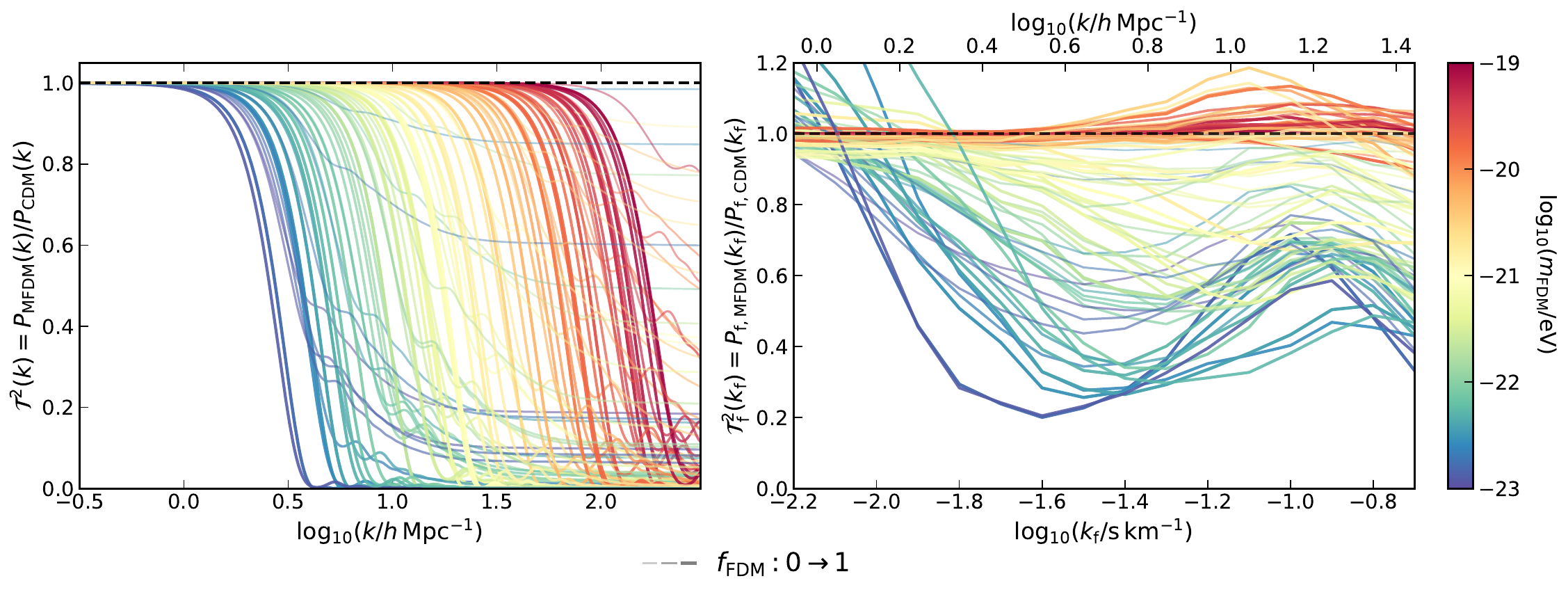}
    \caption{
    Squared transfer functions for the MFDM models used in the training set.
    The left panel shows the squared transfer function of the linear matter power spectrum, $\mathcal{T}^2(k)=P_{\mathrm{MFDM}}(k)/P_{\mathrm{CDM}}(k)$, at the initial redshift $z=99$.
    The right panel shows the corresponding squared transfer function of the 1D flux power spectrum, $\mathcal{T}_{\mathrm{f}}^2(k_{\mathrm{f}})=P_{\mathrm{f,MFDM}}(k_{\mathrm{f}})/P_{\mathrm{f,CDM}}(k_{\mathrm{f}})$, at $z=5.0$.
    The 1D flux power spectra are computed from paired MFDM and CDM simulations with the same input thermal history and are rescaled to the same effective optical depth.
    The colors of the curves indicate different values of $\log_{10}(m_{\mathrm{FDM}}/\mathrm{eV})$, while the line width increases with $f_{\mathrm{FDM}}$ from 0 to 1.
    }
    \label{fig:mfdm_transfer}
\end{figure}

Figure~\ref{fig:mfdm_transfer} shows the squared transfer functions for the 100 training pairs.
The left panel shows the squared transfer function of the linear matter power spectrum, $\mathcal{T}^2(k)$, at the initial redshift $z=99$. 
The right panel shows the corresponding squared transfer function of the 1D flux power spectrum, $\mathcal{T}_{\mathrm{f}}^2(k_{\mathrm{f}})$, at $z=5.0$, computed after rescaling the paired MFDM and CDM 1D flux power spectra to the same effective optical depth.

We note that a small fraction of the models shows an enhancement rather than a suppression of the 1D flux power spectrum on small scales in the right panel of Figure~\ref{fig:mfdm_transfer}.
A qualitatively similar feature can also be seen in ref.~\cite{2019MNRAS.482.3227N}. 
This can arise because non-linear gravitational evolution can partly restore the initially suppressed power on the relevant scales toward the unsuppressed CDM level~\cite{2018MNRAS.478.3935N}. 
Moreover, the 1D flux power spectrum is measured in velocity space and depends not only on the gas density distribution, but also on the gas temperature, pressure smoothing, and peculiar velocities.
These gas properties and dynamical effects can be modified by non-linear evolution when the initial matter power spectrum is suppressed.
In addition, the 1D flux power spectra shown here are computed after rescaling the optical depths to the same effective optical depth.
Therefore, a suppression of the initial matter power spectrum does not necessarily lead to a suppression of the 1D flux power spectrum at all $k_{\mathrm{f}}$.
We have also verified that this behavior persists when the particle mass is reduced by a factor of eight, suggesting that it is unlikely to be caused by insufficient numerical resolution.

\subsection{Two-stage neural network emulator}\label{sec:3.2}
The thermal parameters $(T_0,\gamma,u_0)$ have clear physical meanings and provide a useful description of how the thermal state of the IGM affects the 1D flux power spectrum. 
However, this three-parameter description may not fully capture the relevant aspects of the thermal history that enter the 1D flux power spectrum, such as the effect of peculiar velocities. 
Therefore, if $(T_0,\gamma,u_0)$ are used directly as part of the emulator input, the emulator accuracy may be limited by this incomplete description.
To avoid this limitation, we instead use the input thermal history parameters of the simulations, $(z_{\mathrm{rei}}, H_{\mathrm{A}},H_{\mathrm{S}})$, as part of the emulator input. 
These parameters directly specify the thermal history in the simulations and are therefore not affected by the possible incompleteness of the $(T_0,\gamma,u_0)$ description. 

A simple and widely used method for constructing an emulator is linear interpolation~\cite{2013PhRvD..88d3502V,2017PhRvD..96b3522I,2024PhRvD.109d3511I}.
However, parameter inference based on linear interpolation can become time-consuming when the parameter space is high-dimensional and the number of data points is large.
In addition, the effects of dark matter and IGM parameters on the 1D flux power spectrum can be highly non-linear and can exhibit complicated degeneracies.
Therefore, when only a finite number of simulations are available, linear interpolation may not provide sufficient accuracy across the full parameter space.
These considerations motivate us to adopt a more flexible and efficient emulator construction method.

Gaussian process emulators provide a flexible non-parametric way to interpolate simulation outputs across parameter space.
They model the 1D flux power spectrum as a stochastic function of the input parameters and provide both a mean prediction and an estimate of the emulator uncertainty~\cite{2019JCAP...02..050B}.
Refs.~\cite{2019JCAP...02..031R,2019JCAP...02..050B,2021PhRvD.103d3526R,2021PhRvL.126g1302R,2022PhRvL.128q1301R} proposed and applied an optimized Gaussian process emulator.
In this approach, an initial Gaussian process emulator is trained on a set of initial simulations, and additional simulations are then added adaptively according to the resulting parameter posterior distribution and emulator uncertainty.
This procedure is repeated until the parameter posterior distribution converges.
For pure or single-component dark matter models, this method can provide a well-converged posterior distribution.
However, in our MFDM case, the additional fraction parameter introduces a strong degeneracy between $m_{\mathrm{FDM}}$ and $f_{\mathrm{FDM}}$.
In particular, smaller values of $m_{\mathrm{FDM}}$ can be allowed if $f_{\mathrm{FDM}}$ is sufficiently small.
This leads to a broad posterior distribution in the $m_{\mathrm{FDM}}$--$f_{\mathrm{FDM}}$ plane.
As a result, an optimized Gaussian process emulator may require a large number of additional simulations to achieve posterior convergence.
In addition, when a large number of data points are used, both training the Gaussian process emulator and evaluating it during parameter inference can be computationally expensive.
This adaptive strategy is also less flexible in practice, since changes in the adopted priors or other inference settings can modify the posterior distribution and may therefore require further simulations.
For these reasons, a Gaussian process emulator is not the optimal choice for our analysis.

Inspired by refs.~\cite{2018PhRvD..98h3540M,2022JCAP...10..032H,2025PhRvD.112d3502G}, we propose to use a two-stage neural network emulator, which can guarantee both accuracy and speed.
We notice that, since MFDM can produce a broad and non-trivial posterior distribution in the $m_{\mathrm{FDM}}-f_{\mathrm{FDM}}$ plane, a broad and well-controlled coverage of this two-dimensional dark matter parameter space is particularly important. 
Compared with a grid that samples only a limited set of mixed dark matter parameters, Latin hypercube sampling provides broader and more uniform coverage of the $m_{\mathrm{FDM}}-f_{\mathrm{FDM}}$ plane.
We further validate the emulator with 10 independent simulations that are not included in the training set, providing a conservative test of the interpolation accuracy in genuinely unseen regions of the MFDM parameter space.
In our case, the dependence of the 1D flux power spectrum on $m_{\mathrm{FDM}}$ becomes very weak when $m_{\mathrm{FDM}}$ is large.
If we directly train an emulator to predict $P_{\mathrm{f,MFDM}}$ from all input parameters, this weak dependence on $m_{\mathrm{FDM}}$ can be easily absorbed by variations in the thermal history parameters.
To mitigate this problem, we adopt a two-stage structure.
The first stage predicts the CDM 1D flux power spectrum, $P_{\mathrm{f,CDM}}(k_{\mathrm{f}})$, for a given redshift, IGM thermal history, and optical depth rescaling parameter.
The second stage predicts the squared transfer function of the 1D flux power spectrum, $\mathcal{T}_{\mathrm{f}}^2(k_{\mathrm{f}})$.

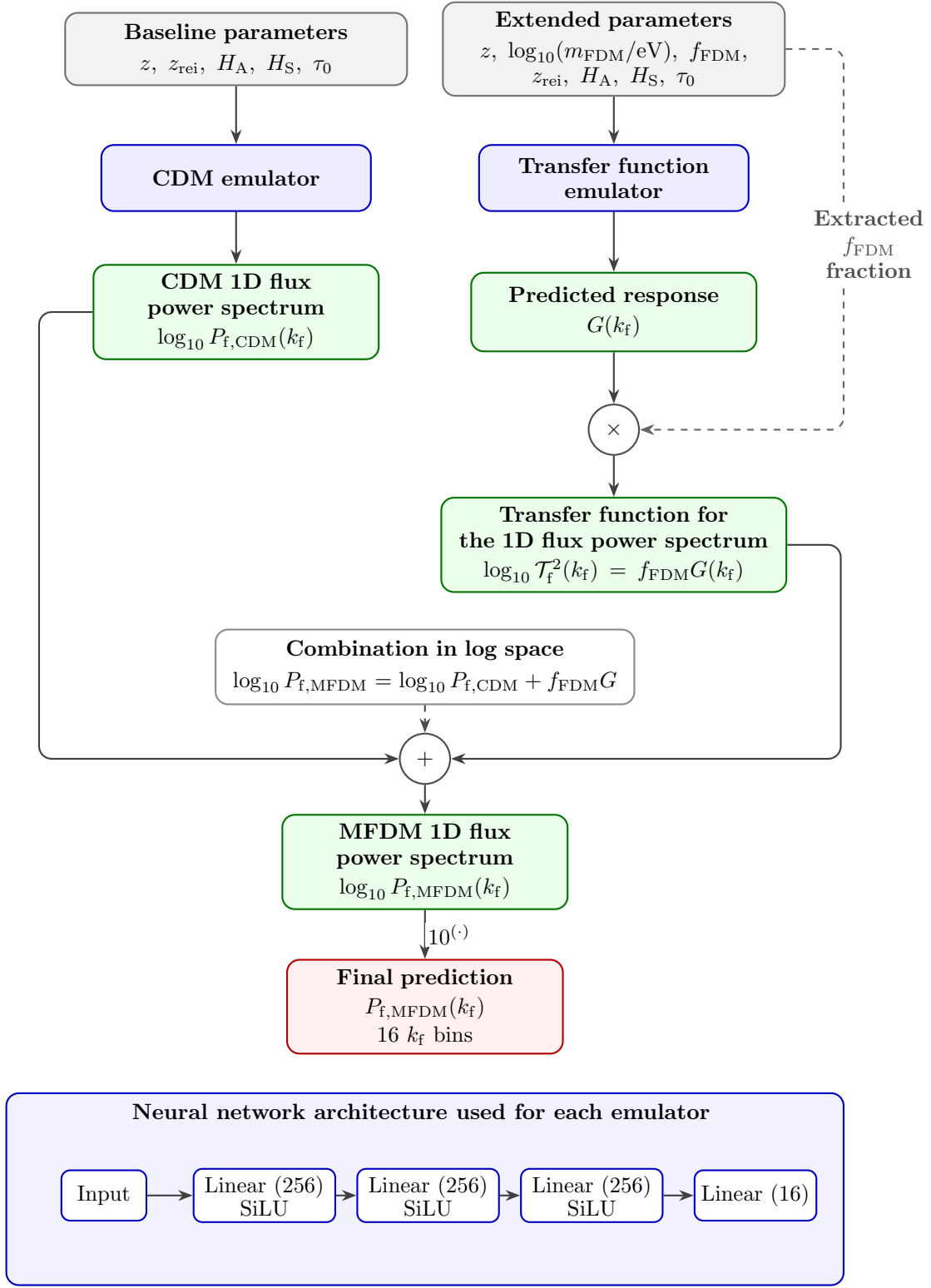
\begin{figure}
\centering
\begin{tikzpicture}[
  >=Stealth,
  font=\normalfont\small,
  line width=0.8pt,
  inputbox/.style={rectangle,rounded corners=2mm,draw=black!55,fill=gray!10,align=center,text width=5.15cm,minimum height=1.15cm,inner sep=4pt},
  emubox/.style={rectangle,rounded corners=2mm,draw=blue!75!black,fill=blue!7,align=center,text width=4.00cm,minimum height=1.05cm,inner sep=4pt},
  specbox/.style={rectangle,rounded corners=2mm,draw=green!45!black,fill=green!8,align=center,text width=4.25cm,minimum height=1.25cm,inner sep=4pt},
  widegreen/.style={rectangle,rounded corners=2mm,draw=green!45!black,fill=green!8,align=center,text width=5.20cm,minimum height=1.30cm,inner sep=4pt},
  finalbox/.style={rectangle,rounded corners=2mm,draw=red!70!black,fill=red!6,align=center,text width=4.10cm,minimum height=1.25cm,inner sep=4pt},
  opbox/.style={circle,draw=black!70,fill=white,minimum size=8.0mm,inner sep=0pt},
  notebox/.style={rectangle,rounded corners=2mm,draw=black!45,fill=white,align=center,text width=6.30cm,minimum height=1.00cm,inner sep=5pt},
  archframe/.style={rectangle,rounded corners=2mm,draw=blue!75!black,fill=blue!5,minimum width=13.30cm,minimum height=3.05cm,inner sep=5pt},
  layerbox/.style={rectangle,rounded corners=1.4mm,draw=blue!75!black,fill=white,align=center,minimum height=0.80cm,inner sep=3pt},
  arr/.style={->,draw=black!75,line width=0.8pt},
  darr/.style={->,draw=black!60,dashed,line width=0.8pt}
]

\node[inputbox] (thermal) at (-3.00, 0.0) {
  \textbf{Baseline parameters}\\[0.7mm]
  $z,\ z_\mathrm{rei},\ H_\mathrm{A},\ H_\mathrm{S},\ \tau_{0}$
};

\node[emubox] (cdmemu) at (-3.00, -2.05) {
  \textbf{CDM emulator}
};

\node[specbox] (cdmout) at (-3.00, -4.18) {
  \textbf{CDM 1D flux}\\[-0.2mm]
  \textbf{power spectrum}\\[0.5mm]
  $\log_{10}P_\mathrm{f,CDM}(k_\mathrm{f})$
};

\node[inputbox] (mfdmin) at (3.00, 0.0) {
  \textbf{Extended parameters}\\[0.7mm]
  $z,\ \log_{10}(m_{\mathrm{FDM}}/\mathrm{eV}),\ f_\mathrm{FDM},$\\[-0.2mm]
  $z_\mathrm{rei},\ H_\mathrm{A},\ H_\mathrm{S},\ \tau_{0}$
};

\node[emubox] (tfemu) at (3.00, -2.05) {
  \textbf{Transfer function}\\[-0.2mm]
  \textbf{emulator}
};

\node[specbox] (gout) at (3.00, -4.18) {
  \textbf{Predicted response}\\[0.5mm]
  $G(k_\mathrm{f})$
};

\node[opbox] (times) at (3.00, -6.05) {$\times$};

\node[widegreen] (tfout) at (3.00, -7.88) {
  \textbf{Transfer function for}\\[-0.2mm]
  \textbf{the 1D flux power spectrum}\\[0.5mm]
  $\log_{10}\mathcal{T}_\mathrm{f}^{2}(k_\mathrm{f})=f_\mathrm{FDM}G(k_\mathrm{f})$
};

\draw[arr] (thermal) -- (cdmemu);
\draw[arr] (cdmemu) -- (cdmout);
\draw[arr] (mfdmin) -- (tfemu);
\draw[arr] (tfemu) -- (gout);
\draw[arr] (gout) -- (times);
\draw[arr] (times) -- (tfout);

\draw[darr, rounded corners=2mm] (mfdmin.east) -- (6.65,0) |- (times.east);
\node[fill=white, text=black!70, align=center, text width=2.05cm, inner sep=2pt]
  at (7.05, -3.10) {\textbf{Extracted}\\$f_\mathrm{FDM}$ \textbf{fraction}};

\node[notebox] (comb) at (0.0, -9.82) {
  \textbf{Combination in log space}\\[0.7mm]
  $\log_{10}P_\mathrm{f,MFDM} = \log_{10}P_\mathrm{f,CDM} + f_\mathrm{FDM}G$
};

\node[opbox] (plus) at (0.0, -11.28) {$+$};

\draw[arr, rounded corners=2mm] (cdmout.west) -- ++(-0.85,0) |- (plus.west);
\draw[arr, rounded corners=2mm] (tfout.east) -- ++(0.85,0) |- (plus.east);
\draw[darr, color=black!70] (comb.south) -- (plus.north);

\node[specbox] (mfdmlog) at (0.0, -12.92) {
  \textbf{MFDM 1D flux}\\[-0.2mm]
  \textbf{power spectrum}\\[0.5mm]
  $\log_{10}P_\mathrm{f,MFDM}(k_\mathrm{f})$
};

\node[finalbox] (finalpred) at (0.0, -15.22) {
  \textbf{Final prediction}\\[0.5mm]
  $P_\mathrm{f,MFDM}(k_\mathrm{f})$\\[0.2mm]
  {16 $k_\mathrm{f}$ bins}
};

\draw[arr] (plus) -- (mfdmlog);
\draw[arr] (mfdmlog) -- node[midway,right,fill=white,inner sep=1pt] {$10^{(\cdot)}$} (finalpred);

\node[archframe] (arch) at (0.0, -18.12) {};

\node[align=center] at (0.0, -16.88) {
  \textbf{Neural network architecture used for each emulator}
};

\node[layerbox,minimum width=1.35cm] (l0) at (-5.10, -18.22) {Input};
\node[layerbox,minimum width=2.25cm] (l1) at (-2.55, -18.22) {Linear (256)\\[-0.5mm] SiLU};
\node[layerbox,minimum width=2.25cm] (l2) at (0.05, -18.22) {Linear (256)\\[-0.5mm] SiLU};
\node[layerbox,minimum width=2.25cm] (l3) at (2.65, -18.22) {Linear (256)\\[-0.5mm] SiLU};
\node[layerbox,minimum width=1.75cm] (l4) at (5.25, -18.22) {Linear (16)};

\draw[arr] (l0) -- (l1);
\draw[arr] (l1) -- (l2);
\draw[arr] (l2) -- (l3);
\draw[arr] (l3) -- (l4);

\end{tikzpicture}
    \caption{
    Structure of the two-stage neural network emulator.
    The first stage emulator predicts the baseline CDM 1D flux power spectrum, $\log_{10}P_{\mathrm{f,CDM}}(k_{\mathrm{f}})$, from the redshift, the input thermal history parameters, and the optical depth rescaling parameter.
    The second stage emulator predicts the MFDM response function, $G(k_{\mathrm{f}})$, from the extended parameter set that includes $\log_{10}(m_{\mathrm{FDM}}/\mathrm{eV})$ and $f_{\mathrm{FDM}}$.
    The two outputs are combined in log space as
    $\log_{10}P_{\mathrm{f,MFDM}}=\log_{10}P_{\mathrm{f,CDM}}+f_{\mathrm{FDM}}G$,
    which enforces the correct CDM limit when $f_{\mathrm{FDM}}=0$.
    Both emulators use the same fully connected feed-forward neural network architecture, with three hidden layers of 256 neurons and SiLU activation functions.
    The output layer has 16 neurons, corresponding to the 16 $k_{\mathrm{f}}$ bins.
    }
    \label{fig:emulator_structure}
\end{figure}

The structure of our emulator is shown in Figure~\ref{fig:emulator_structure}.
The first stage emulator takes the baseline parameters, $(z,z_{\mathrm{rei}},H_{\mathrm{A}},H_{\mathrm{S}},\tau_0)$, as input and predicts the CDM 1D flux power spectrum in log space, $\log_{10} P_{\mathrm{f,CDM}}(k_{\mathrm{f}})$.
Here $z$ denotes the redshift of the simulation snapshot, and the output is a vector with 16 components corresponding to the 16 $k_{\mathrm{f}}$ bins.
Compared with the first stage emulator, the second stage emulator additionally takes two dark matter parameters, 
$\log_{10}(m_{\mathrm{FDM}}/\mathrm{eV})$ and $f_{\mathrm{FDM}}$, 
and predicts the response function $G(k_{\mathrm{f}})$, which is defined as
\begin{equation}
    G(k_{\mathrm{f}})
    =
    \frac{
    \log_{10} P_{\mathrm{f,MFDM}}(k_{\mathrm{f}})
    -
    \log_{10} P_{\mathrm{f,CDM}}(k_{\mathrm{f}})
    }{
    f_{\mathrm{FDM}}
    },
    \quad \mathrm{when}\ f_{\mathrm{FDM}}>0.
\end{equation}
For $f_{\mathrm{FDM}}=0$, we set the MFDM contribution to zero.
The final prediction is then obtained by combining the two emulators in log space,
\begin{equation}
\label{eq:emulator_combination}
    \log_{10} P_{\mathrm{f,MFDM}}(k_{\mathrm{f}})
    =
    \log_{10} P_{\mathrm{f,CDM}}(k_{\mathrm{f}})
    +
    f_{\mathrm{FDM}}G(k_{\mathrm{f}}).
\end{equation}
Equivalently, this can be written as
\begin{equation}
    \log_{10}\mathcal{T}_{\mathrm{f}}^{2}(k_{\mathrm{f}})
    =
    f_{\mathrm{FDM}}G(k_{\mathrm{f}}).
\end{equation}
This construction enforces the correct CDM limit when $f_{\mathrm{FDM}}=0$.
It also helps isolate the dark matter signal from the baseline CDM prediction.
Both emulators use the same fully connected feed-forward neural network architecture.
The network consists of three hidden layers, each with 256 neurons.
We apply the SiLU activation function after each hidden layer~\cite{2017arXiv170203118E}.
The output layer has 16 neurons, corresponding to the 16 $k_{\mathrm{f}}$ bins.
All input parameters are rescaled with a min-max scaler before training.
Each emulator stage is trained on $3\times 31\times N_{\mathrm{pair}}$ data vectors.
These correspond to three redshift snapshots, 31 values of $\tau_0$, and $N_{\mathrm{pair}}$ paired MFDM and CDM simulation points.

For each $N_{\mathrm{pair}}$, we train five models using a five-fold split over the simulation indices.
All redshifts and all values of $\tau_0$ from the same simulation are kept in the same fold.
The final emulator prediction is taken as the average of the five trained models.
For the first stage emulator, we minimize a weighted mean-squared error, with weights set by the observational uncertainties converted to log space.
For the second stage emulator, we do not optimize the error on $G(k_{\mathrm{f}})$ itself.
Instead, we optimize the residual term, $f_{\mathrm{FDM}}G(k_{\mathrm{f}})$, using the same weighted mean-squared error.
This choice makes the training objective more directly related to the quantity entering the likelihood.
{It also preserves the correct CDM limit by forcing the MFDM contribution to vanish when $f_{\mathrm{FDM}}=0$, and improves the accuracy of the predicted MFDM residual in the small-$f_{\mathrm{FDM}}$ regime.}
We train the networks using the AdamW optimizer~\cite{2017arXiv171105101L}.
The initial learning rate is $10^{-3}$, and the weight decay is set to $10^{-6}$.
The learning rate is reduced when the validation loss stops improving, and early stopping is used to avoid overfitting.

\begin{figure}
    \centering
    \includegraphics[width=\linewidth]{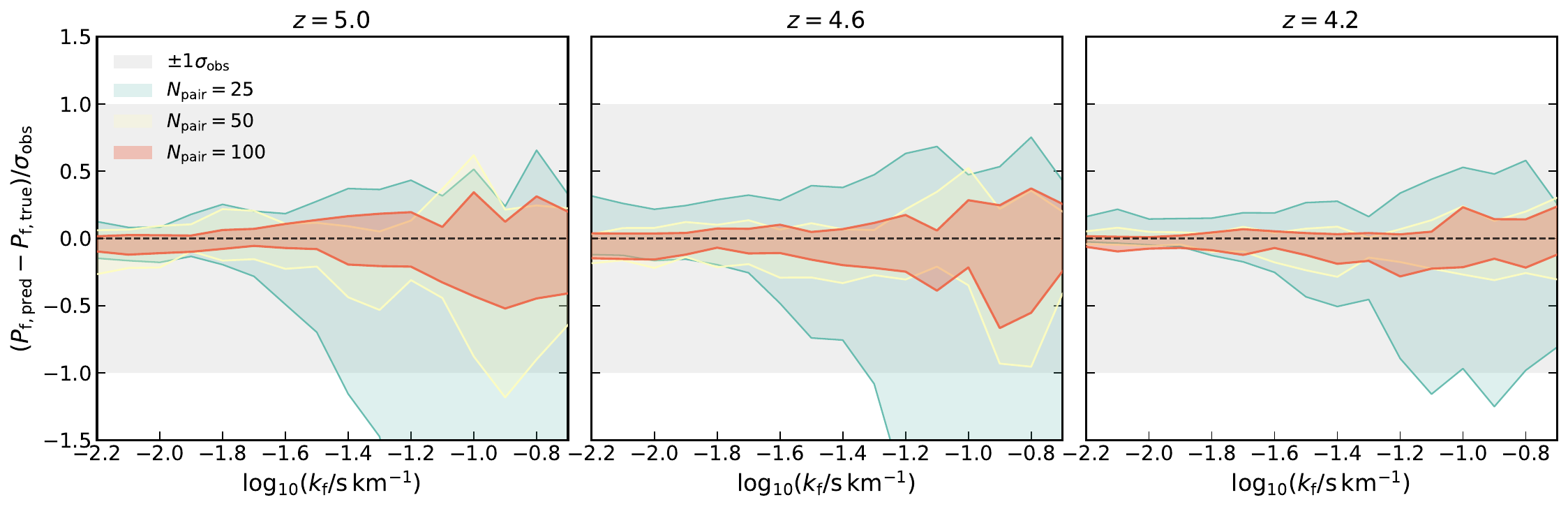}
    \caption{
    Emulator validation using 10 independent MFDM simulations that are not included in the training set.
    The vertical axis shows the emulator error normalized by the observational uncertainty, {$(P_{\mathrm{f,pred}} - P_{\mathrm{f,true}})/\sigma_{\mathrm{obs}}$}.
    The three panels correspond to $z=5.0$, 4.6, and 4.2.
    The colored shaded regions show the 16th to 84th percentiles of the error distribution for training sets with $N_{\mathrm{pair}}=25$, 50, and 100 paired simulations.
    The grey band indicates the observational $1\sigma$ uncertainty.
    The emulator accuracy improves as $N_{\mathrm{pair}}$ increases.
    The case with $N_{\mathrm{pair}}=50$ already gives good performance, while $N_{\mathrm{pair}}=100$ further reduces the emulator error over the full range of scales used in our analysis.
    }
    \label{fig:emulator_error}
\end{figure}

We validate the emulator using 10 independent MFDM simulations that are not used in the training process.
{Figure~\ref{fig:emulator_error} shows the emulator error normalized by the observational uncertainty,
$(P_{\mathrm{f,pred}} - P_{\mathrm{f,true}})/\sigma_{\mathrm{obs}}$,
where $P_{\mathrm{f,pred}}$ is the 1D flux power spectrum predicted by the emulator, $P_{\mathrm{f,true}}$ is the corresponding result measured directly from the simulation, and $\sigma_{\mathrm{obs}}$ denotes the observational uncertainty.}
As the number of paired simulations increases, the emulator error decreases significantly.
For $N_{\mathrm{pair}}=50$, the emulator already reaches a good accuracy at all three redshifts.
Increasing the training set to $N_{\mathrm{pair}}=100$ further reduces the error and gives a more stable prediction.
We can find that the largest errors appear at the smallest scales, where the 1D flux power spectrum is more sensitive to the detailed non-linear evolution.
Even at these scales, the 68\% error region remains smaller than the current observational $1\sigma$ uncertainty.
We therefore use the emulator trained with $N_{\mathrm{pair}}=100$ in our final analysis.
{In addition, the emulator error is not very sensitive to the detailed architecture or training choices of our neural network emulator.
We have tested several choices, including the number of layers, the number of neurons per layer, the activation function, and the number of training folds.
We find that, as long as the two-stage structure of our neural network emulator is kept fixed, the resulting emulator errors change only mildly.
We therefore adopt the configuration with the best overall validation performance.}
Further discussion on the validity of the emulator is provided in appendix~\ref{A}.

\section{Constraints and results} \label{sec:4}
To derive constraints on the dark matter and IGM parameters, we adopt a Gaussian likelihood function and perform Bayesian parameter inference with \texttt{emcee}~\cite{2013PASP..125..306F}, using the trained emulator. 
We allow $(T_0,\gamma,u_0,\tau_0)$ to vary independently at different redshifts, which gives 13 free parameters for the PFDM case and 14 free parameters for the MFDM case.
Note that the physically interpretable parameters describing the thermal state of the IGM are $(T_0,\gamma,u_0)$, but our emulator is built in terms of the simulation input parameters $(z_{\mathrm{rei}}, H_{\mathrm{A}}, H_{\mathrm{S}})$.
We therefore construct a simple radial basis function interpolator~\cite{Buhmann_Dyn_1993,2021PhRvD.103d3526R}  to obtain an accurate mapping from $(z_{\mathrm{rei}}, H_{\mathrm{A}}, H_{\mathrm{S}})$ to $(T_0,\gamma,u_0)$ at each redshift.
An advantage of this mapping is that it naturally excludes unphysical combinations of the thermal parameters, such as models with a very high $T_0$ but a very low $u_0$.

In the parameter inference, we adopt priors that are uniform in 
$\log_{10}(m_{\mathrm{FDM}}/\mathrm{eV})$, $f_{\mathrm{FDM}}$, and $\tau_0$.
For the IGM thermal parameters, we sample uniformly over $(z_{\mathrm{rei}}, H_{\mathrm{A}}, H_{\mathrm{S}})$ at each redshift and then map the sampled points to $(T_0,\gamma,u_0)$.
The prior ranges of these sampled parameters are the same as the parameter ranges described in section~\ref{sec:3.1}.
This procedure may introduce a prior-volume effect in the derived thermal parameters, but as we show below that this effect does not affect our conclusions.
At the same time, we impose a weak continuity prior by requiring that the variations of $T_0$ and $u_0$ between adjacent redshift bins do not exceed $5000~\mathrm{K}$ and $5~\mathrm{eV}\,m_{\mathrm{p}}^{-1}$, respectively~\cite{2021PhRvL.126g1302R}.
In addition, following ref.~\cite{2024PhRvD.109d3511I}, we impose Gaussian priors on $T_0$ at each redshift.
The prior mean values at $z=5.0$, $4.6$, and $4.2$ are $9286.5$, $8986.5$, and $9155.5~\mathrm{K}$, respectively, each with an uncertainty of $1000~\mathrm{K}$.

\subsection{Constraints on pure FDM}\label{sec:4.1}

\begin{figure}
    \centering
    \includegraphics[width=0.7\linewidth]{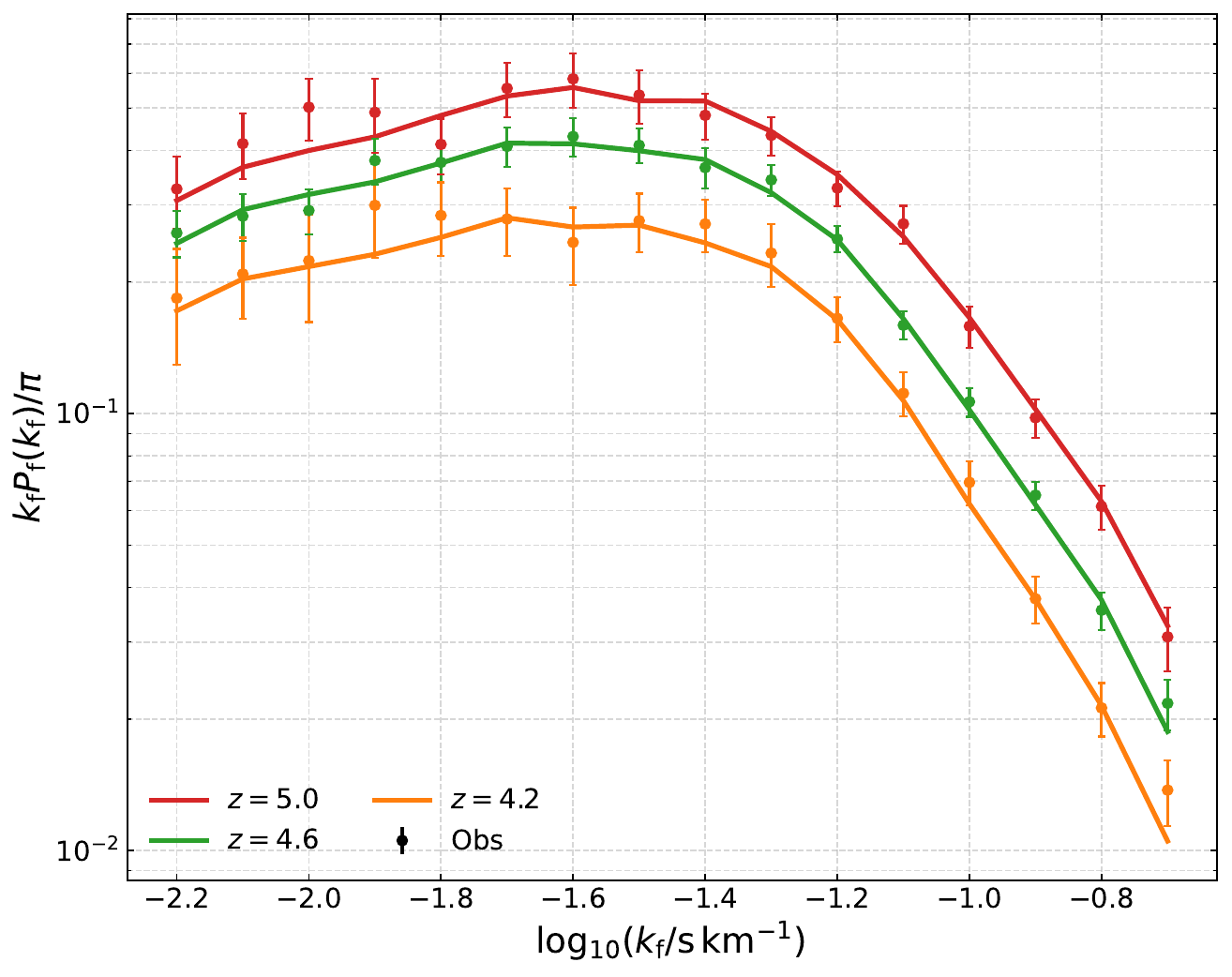}
    \caption{
    Comparison between the best-fitting PFDM model and the observed 1D flux power spectrum.
    The data points with error bars show the observed 1D flux power spectra with $1\sigma$ errors at $z=5.0$, $4.6$, and $4.2$, and the solid curves show the corresponding best-fitting model predictions.
    }
    \label{fig:bestfit_pure_fdm}
\end{figure}

\begin{figure}
    \centering
    \includegraphics[width=\linewidth]{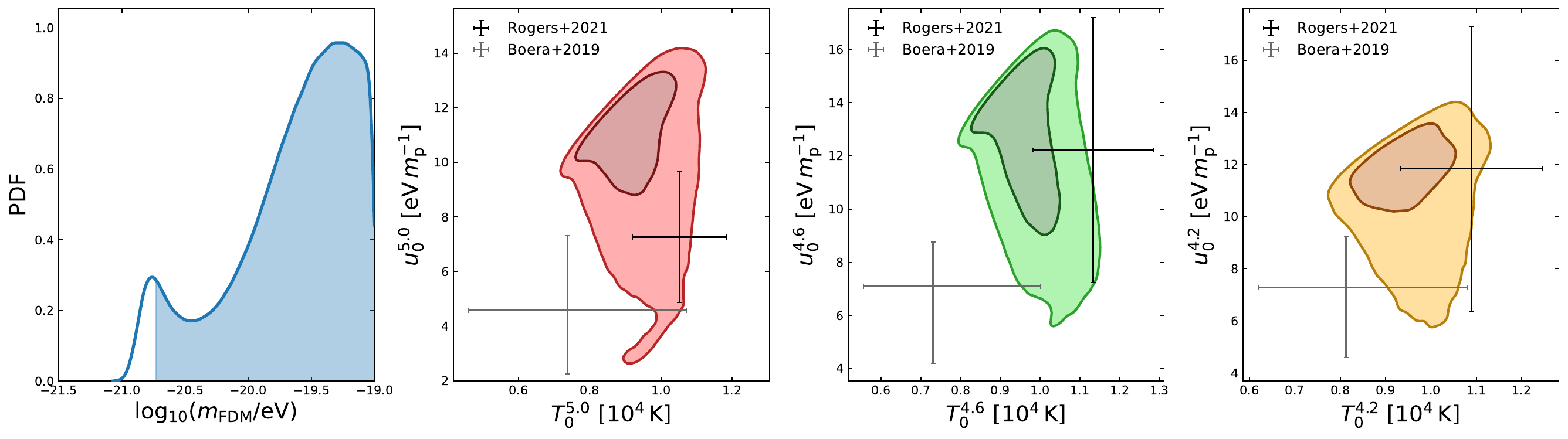}
    \caption{
    The PDF of $\log_{10}(m_{\mathrm{FDM}}/\mathrm{eV})$ and contour maps of the thermal parameters in the PFDM model.
    The contour maps show the 68\% and 95\% credible levels in the $T_0-u_0$ plane at $z=5.0$, $4.6$, and $4.2$.
    The shaded region in the 1D PDF indicates the parameter region allowed by the 95\% lower limit on $m_{\mathrm{FDM}}$.
    For comparison, we also show the 95\% thermal constraints from Rogers+2021~\cite{2021PhRvL.126g1302R} and Boera+2019~\cite{2019ApJ...872..101B}.
    }
    \label{fig:pure_fdm_posterior}
\end{figure}

In Figure~\ref{fig:bestfit_pure_fdm}, we compare the best-fitting PFDM model with the observed 1D flux power spectrum, and we can find that the best-fitting model provides a good fit to the data.
Figure~\ref{fig:pure_fdm_posterior} shows the probability density function (PDF) of $\log_{10}(m_{\mathrm{FDM}}/\mathrm{eV})$ and the contour maps of the thermal parameters in the PFDM analysis.
For comparison, we also show the thermal parameter constraints from refs.~\cite{2019ApJ...872..101B,2021PhRvL.126g1302R}, and we can see that our constraints on the thermal parameters are generally consistent with the previous results.
After marginalizing over the IGM parameters, we obtain the following 95\% lower limit:
\begin{equation}
\label{purefdm_con}
    m_{\mathrm{FDM}} > 1.9 \times 10^{-21}~\mathrm{eV}.
\end{equation}

Our result is consistent with no strong preference for PFDM over CDM, but the lower bound is about one order of magnitude weaker than that of ref.~\cite{2021PhRvL.126g1302R}, which reported $m_{\mathrm{FDM}} > 2.0 \times 10^{-20}~\mathrm{eV}$.
This difference may be due to two main reasons.
First, ref.~\cite{2021PhRvL.126g1302R} used glass initial conditions for the gas component, while our simulations initialize dark matter and gas particles on offset grids~\cite{2024MNRAS.530.4920K}.
Second, the emulator construction is different.
As discussed in section~\ref{sec:3.2}, ref.~\cite{2021PhRvL.126g1302R} used an optimized Gaussian process emulator for a PFDM model, which allowed them to obtain a well-converged and stringent constraint for this single-component dark matter scenario.
In contrast, our two-stage neural network emulator is constructed to cover the broader MFDM parameter space while maintaining sufficient accuracy, with both $m_{\mathrm{FDM}}$ and $f_{\mathrm{FDM}}$ varied simultaneously.
Therefore, our approach is more general and is designed to handle the broad and non-trivial posterior distribution of MFDM, although it may be less optimized for the narrow PFDM posterior than a dedicated Gaussian process emulator.

We also note that the PDF of $\log_{10}(m_{\mathrm{FDM}}/\mathrm{eV})$ shows a small secondary feature around $-20.8$.
This feature appears because the suppression produced by $\log_{10}(m_{\mathrm{FDM}}/\mathrm{eV})\simeq -20.8$ can be partly compensated by a lower value of $u_0$.
Similar behaviour has also been reported in other Lyman-$\alpha$ forest analyses, including the warm dark matter analysis of ref.~\cite{2023PhRvD.108b3502V}, which used the same data set, and studies based on different data sets~\cite{2017PhRvD..96b3522I,2017PhRvL.119c1302I,2017PhLB..773..258G}.
Apart from this small feature, we do not find a clear degeneracy between $\log_{10}(m_{\mathrm{FDM}}/\mathrm{eV})$ and the thermal parameters $u_0$ or $T_0$.
A more detailed view of the posterior distributions for the parameters of interest is presented in appendix~\ref{B}.

Introducing an additional prior on the Thomson optical depth may help break the mild degeneracy between  $m_{\mathrm{FDM}}$ and low $u_0$, for example in the spirit of the method discussed in ref.~\cite{2025arXiv251000107G}.
However, as discussed in section~\ref{sec:3.1}, our hydrodynamical simulations assume photoionization equilibrium and do not solve the full non-equilibrium equations.
Therefore, the Thomson optical depth computed from our UV background should not be compared directly with the physical value.
In addition, our construction of the UV background differs from that of ref.~\cite{2025arXiv251000107G}, and the mapping between the IGM thermal parameters and the Thomson optical depth depends on the details of the UV background implementation.
We therefore leave a self-consistent inclusion of the Thomson optical depth prior to future work.
Nevertheless, if we interpret our inferred IGM thermal parameters using the mapping of ref.~\cite{2025arXiv251000107G}, the corresponding Thomson optical depth is generally consistent with the Planck 2018 value~\cite{2020A&A...641A...6P}.
This agreement provides a useful consistency check for our parameter inference results.

We also note that Bayesian parameter inference can be affected by prior-volume effects.
This may influence the inferred lower bound on $m_{\mathrm{FDM}}$.
As a robustness check, we therefore construct a frequentist 95\% confidence interval using the Neyman construction with a profile-likelihood test statistic~\cite{2025PhRvD.111h3504H,2026arXiv260325731S}.
This gives $m_{\mathrm{FDM}} > 1.1 \times 10^{-21}~\mathrm{eV}$, which is consistent with our Bayesian inference result, suggesting that our constraint is not primarily driven by prior-volume effects.
Finally, ref.~\cite{2024PhRvD.109d3511I} pointed out that the smallest-scale data points used in current high-redshift Lyman-$\alpha$ forest analyses may be less secure.
{We therefore repeat the analysis after removing one, two, or three smallest-scale data points at each redshift.
We find that our final conclusions remain similar to the baseline result, except that the degeneracy around $\log_{10}(m_{\mathrm{FDM}}/\mathrm{eV})\simeq -20.8$ and a lower value of $u_0$ becomes more apparent.
This is expected, since $m_{\mathrm{FDM}}$ and $u_0$ affect the flux power spectrum differently on small scales, and the smallest-scale data points help to partially break this degeneracy.}

\subsection{Constraints on mixed FDM}\label{sec:4.2}

For the MFDM model, it is a natural extension of PFDM and can help alleviate the tension between PFDM and several observations~\cite{2019PhRvD..99f3509L,2019MNRAS.482.3227N}.
We follow the same inference procedure as in the PFDM case, but now include the FDM fraction, $f_{\mathrm{FDM}}$, as an additional free parameter.
As expected, the MFDM constraint shows a strong degeneracy between $m_{\mathrm{FDM}}$ and $f_{\mathrm{FDM}}$, since lighter FDM masses can remain allowed only when the FDM fraction is sufficiently small.

\begin{figure}
    \centering
    \includegraphics[width=0.7\linewidth]{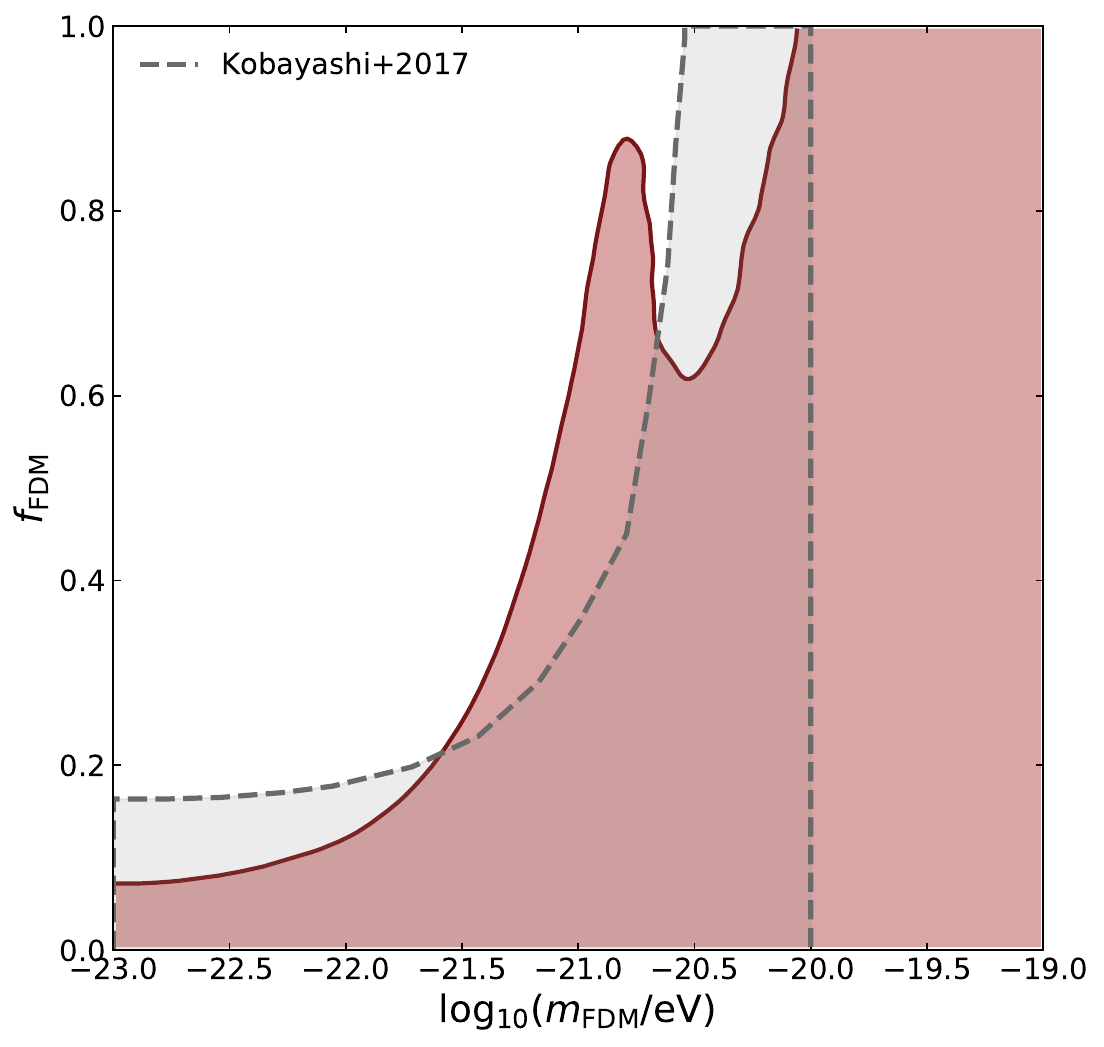}
    \caption{
    The constraint result (red region) of the MFDM parameters, i.e., $\log_{10}(m_{\mathrm{FDM}}/\mathrm{eV})$ vs. $f_{\mathrm{FDM}}$.
    The shaded region corresponds to the parameter space allowed at the 95\% credible level.
    The grey region shows the constraint result from Kobayashi+2017~\cite{2017PhRvD..96l3514K} for comparison.
    }
    \label{fig:mixed_fdm_constraint}
\end{figure}

In Figure~\ref{fig:mixed_fdm_constraint},  we show the constraint result (red region) of the MFDM parameters, i.e., $\log_{10}(m_{\mathrm{FDM}}/\mathrm{eV})$ and $f_{\mathrm{FDM}}$, and the result from ref.~\cite{2017PhRvD..96l3514K} (grey region) is also shown for comparison.
For the 95\% upper limits, we obtain
\begin{equation}
\label{eq:mixed_fdm_con}
    f_{\mathrm{FDM}}
    <
    0.07,\ 0.12,\ 0.65
\end{equation}
for
\begin{equation}
    \log_{10}(m_{\mathrm{FDM}}/\mathrm{eV})
    =
    -23.0,\ -22.0,\ -21.0,
\end{equation}
respectively.
For $\log_{10}(m_{\mathrm{FDM}}/\mathrm{eV}) \gtrsim -20$, the current data do not provide an effective upper limit on $f_{\mathrm{FDM}}$ within the parameter range explored by our emulator, and the PFDM case with $f_{\mathrm{FDM}}=1$ is still consistent with the data.
{As discussed in section~\ref{sec:2.3}, these MFDM limits may be conservative for small $m_{\mathrm{FDM}}$, without including the full Schr\"odinger--Poisson evolution.}

The constraint result can be understood from the limiting behavior of the MFDM transfer function.
For very small $m_{\mathrm{FDM}}$, even a small FDM fraction produces a strong suppression of small-scale power, and therefore $f_{\mathrm{FDM}}$ is tightly constrained.
As $m_{\mathrm{FDM}}$ increases, the suppression moves to smaller scales and becomes difficult to distinguish from the effects of the IGM thermal history, and then the upper limit on $f_{\mathrm{FDM}}$ becomes weaker.
At sufficiently large $m_{\mathrm{FDM}}$, the model approaches the CDM prediction on the scales probed by the current Lyman-$\alpha$ forest data, and the fraction becomes essentially unconstrained.

Compared with ref.~\cite{2017PhRvD..96l3514K}, our constraints are generally stronger, although we adopt a broader prior range in $\log_{10}(m_{\mathrm{FDM}}/\mathrm{eV})$.
The overall improvement can be attributed to several factors.
First, we use higher-resolution observational data that extend to smaller scales, with a highest wavenumber of $k_{\mathrm{f}}\simeq 0.2~\mathrm{s}\,\mathrm{km}^{-1}$.
Second, we adopt a more flexible parameterization of the 1D flux power spectrum.
Third, we perform a more complete exploration of the parameter space and combine it with a two-stage neural network emulator, which provides accurate predictions for the 1D flux power spectrum without relying on linear interpolation or extrapolation.

{Compared with the PFDM case, prior-volume effects can be more important for MFDM.
We therefore also perform the Neyman construction using a profile-likelihood test statistic, and obtain
$f_{\mathrm{FDM}}<0.08,\ 0.17,\ 0.90$
for
$\log_{10}(m_{\mathrm{FDM}}/\mathrm{eV})=-23.0,\ -22.0,\ -21.0$,
respectively.
For $\log_{10}(m_{\mathrm{FDM}}/\mathrm{eV})>-20$, the data do not provide an effective upper limit on $f_{\mathrm{FDM}}$.
The frequentist limits are slightly weaker than the marginalized Bayesian limits, which can be understood as a consequence of prior-volume effects: in the Bayesian analysis, there is a large allowed parameter volume at $\log_{10}(m_{\mathrm{FDM}}/\mathrm{eV})>-20$, where $f_{\mathrm{FDM}}$ remains only weakly constrained.
In addition, to assess the impact of the observational data at the smallest scales, we also repeat the analysis after removing one, two, or three smallest-scale data points at each redshift.
The resulting constraints remain similar to our baseline result, except that the degeneracy around $\log_{10}(m_{\mathrm{FDM}}/\mathrm{eV})\simeq -20.8$ and a lower value of $u_0$ becomes more apparent.
}

\section{Summary and outlook} \label{sec:5}
In this work, we constrain the PFDM and MFDM models with the measurements of the 1D flux power spectrum of the Lyman-$\alpha$ forest at high redshifts, by adopting a dedicated two-stage neural network emulator trained on cosmological hydrodynamical simulations.
The emulator is trained using 100 matched pairs of CDM and MFDM simulations.
The first stage predicts the CDM 1D flux power spectrum, while the second stage predicts the effect of MFDM relative to the CDM baseline.
We show that the emulator is sufficiently accurate compared to the current observational uncertainties.

We then use this emulator to perform Bayesian parameter inference and derive the 95\% limits on the FDM parameters.
For the PFDM model, we find
$m_{\mathrm{FDM}} > 1.9 \times 10^{-21}~\mathrm{eV}$, and 
for the MFDM model, we obtain
$f_{\mathrm{FDM}}<0.07,\ 0.12,\ 0.65$
for
$\log_{10}(m_{\mathrm{FDM}}/\mathrm{eV})=-23.0,\ -22.0,\ -21.0$,
respectively.
For $\log_{10}(m_{\mathrm{FDM}}/\mathrm{eV})\gtrsim -20$, the current data do not provide an effective upper limit on $f_{\mathrm{FDM}}$.
We further test the robustness of our results using a frequentist Neyman construction with a profile-likelihood test statistic and by repeating the analysis after removing the less secure smallest-scale data points.
These tests lead to conclusions consistent with our baseline Bayesian analysis.

Based on our analysis, the results disfavor the simple single-field PFDM scenario, in which all of the dark matter is composed of an ultralight scalar field with a canonical mass around $10^{-22}~\mathrm{eV}$.
Allowing FDM to have an attractive self-interaction may help alleviate this tension~\cite{2023MNRAS.521.2608M}, but this possibility requires further investigation with dedicated numerical simulations.
For the MFDM study, a more complete treatment including the full Schr\"odinger--Poisson evolution may be needed when using the measurements of Lyman-$\alpha$ forest, and external priors or joint analyses with other probes may help reduce or break possible degeneracies between model parameters~\cite{2021MNRAS.506.5848E,2025arXiv251000107G}.
For the emulator developed in this work, we notice that our emulator framework is not limited to the analysis of the MFDM model, and it can be naturally extended to other scenarios that modify the small-scale matter power spectrum, such as primordial magnetic fields and other non-cold dark matter models~\cite{2024PhRvD.109d3511I,2025PhRvL.135g1001P}.
Besides, further improvements to the emulator may also be possible by using more flexible architectures, such as mixture density networks~\cite{Bishop1994MixtureDN}, which can be studied in the future work.

\acknowledgments
Some of the main code, data, and trained emulators used in this work are available at \url{https://github.com/jianxiangl-astro/lya-mfdm}, together with additional reference results that are not shown explicitly in this paper.
We thank Jose O\~norbe, Simeon Bird, Vid Ir\v{s}i\v{c}, Keir Rogers, James Bolton, and Olga Garcia-Gallego for helpful comments, suggestions, and discussions.
J.X.L. and Y.G. acknowledge the support from the CAS Project for Young Scientists in Basic Research (No. YSBR-92), and National Key R\&D Program of China grant Nos. 2022YFF0503404 and 2020SKA0110402. 
This work is also supported by science research grants from the China Manned Space Project with grant Nos. CMS-CSST-2025-A02, CMS-CSST-2021-B01, and CMS-CSST-2021-A01.

\appendix
\section{Emulator validation and mock recovery} \label{A}
\begin{figure}
    \centering
    \includegraphics[width=\linewidth]{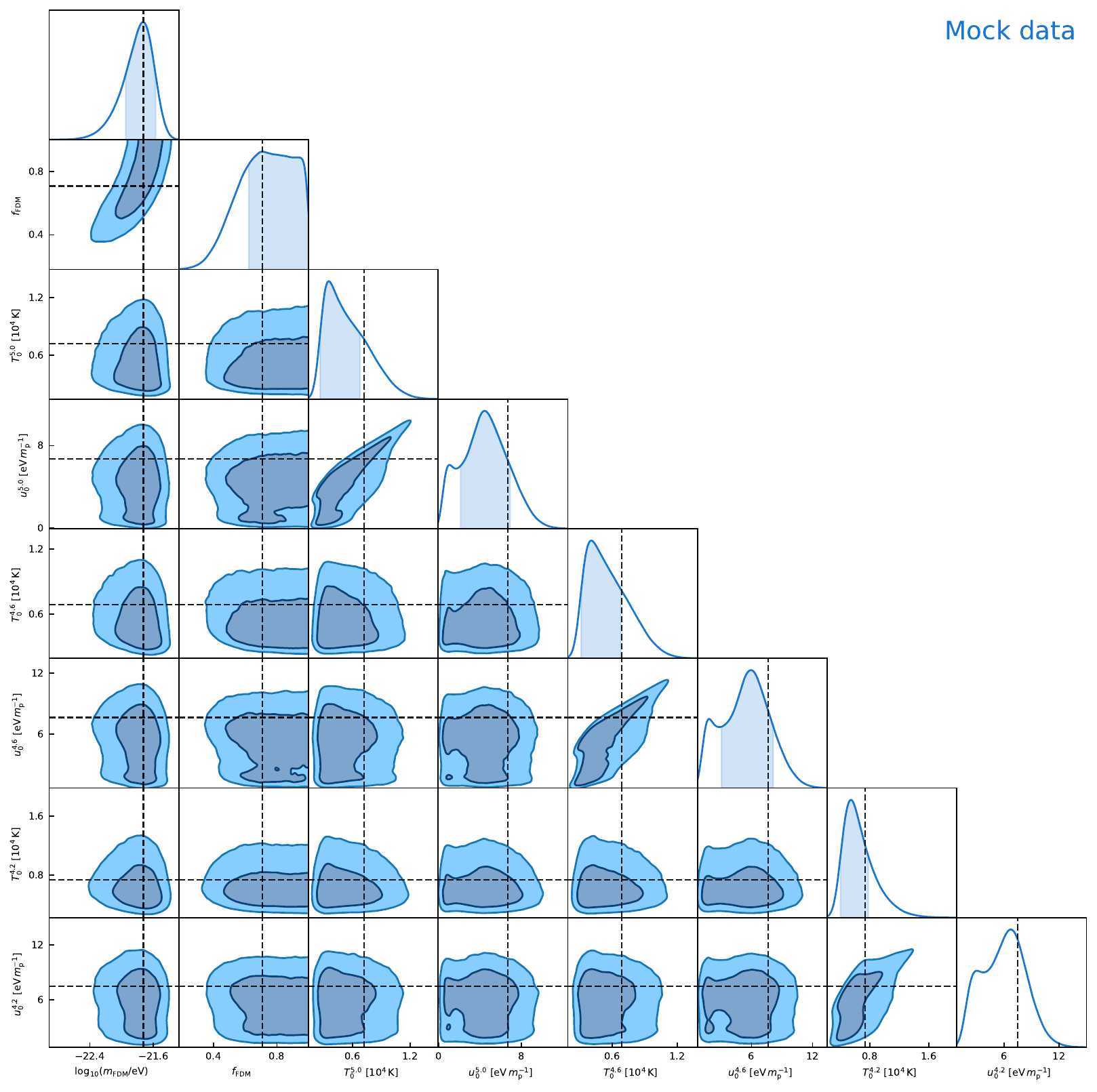}
    \caption{
    The PDFs and contour maps of the parameters recovered from a representative mock MFDM analysis.
    The contours correspond to the 68\% and 95\% credible levels.
    The shaded regions in the 1D PDFs correspond to the 68\% credible intervals.
    The dashed lines indicate the true values.
    }
    \label{fig:mock_recovery}
\end{figure}

We further test the emulator described in section~\ref{sec:3.2} by performing mock recovery analyses.
We use the 10 independent MFDM simulations that are not included in the training set as mock data sets.
For each mock data set, we apply the same inference pipeline as used for the real data.
The only difference is that neither the temperature prior nor the weak continuity prior is imposed.
We then check whether the true values can be recovered, and find that the true values are generally recovered in these tests.
One representative example is shown in Figure~\ref{fig:mock_recovery}.
Although the parameter space is high-dimensional and the degeneracies between the dark matter and IGM thermal parameters, as well as those between the two dark matter parameters themselves, are non-trivial, the posterior distributions are consistent with the true values.
This indicates that the emulator does not introduce a significant bias in these mock recovery tests.

\section{Posterior distributions for the parameters of interest} \label{B}
\begin{figure}
    \centering
    \includegraphics[width=\linewidth]{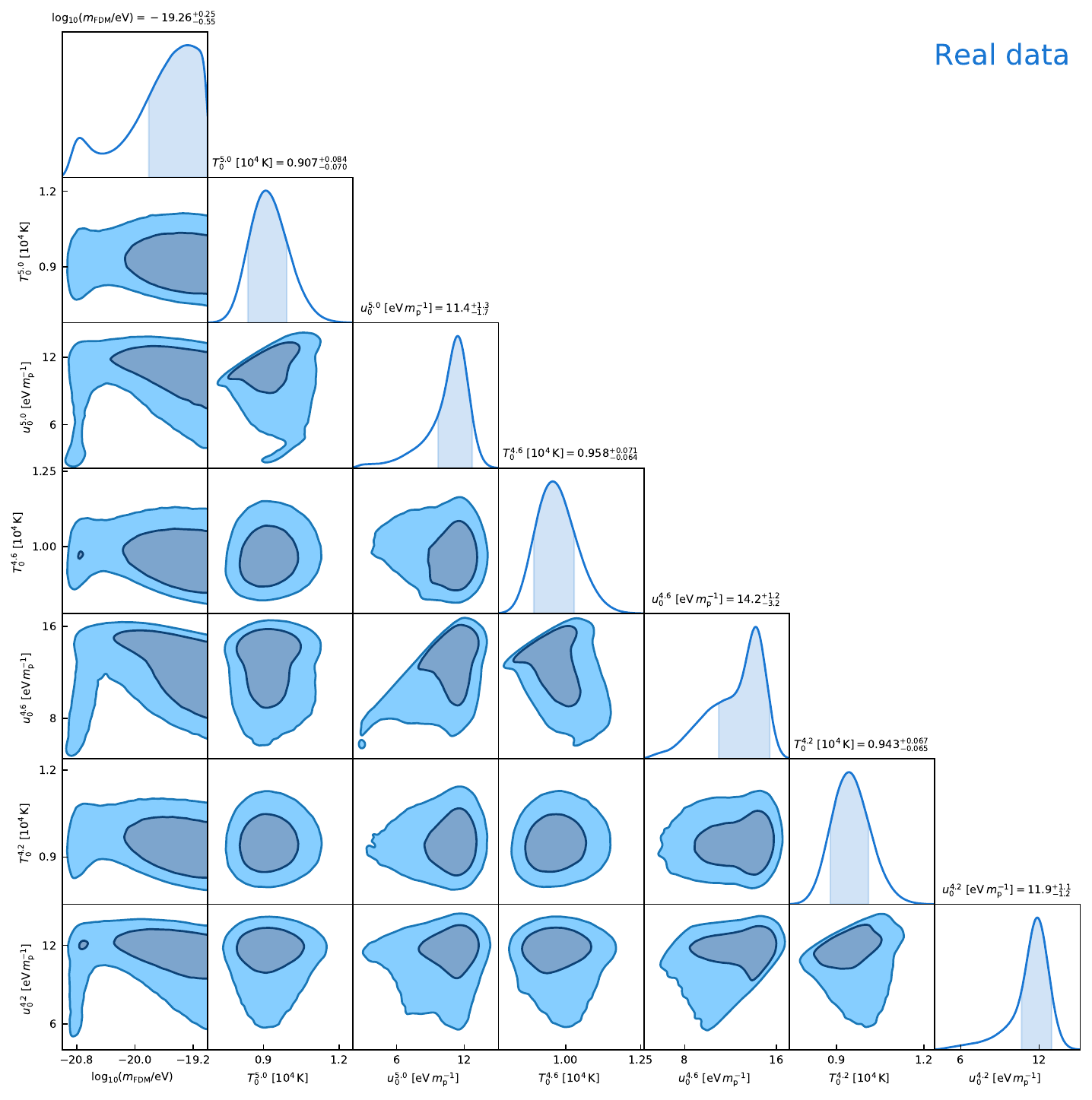}
    \caption{
    The PDFs and contour maps of the parameters of interest in the baseline PFDM analysis.
    These parameters include the FDM mass and the IGM thermal parameters, $T_0$ and $u_0$, at each redshift.
    The contours correspond to the 68\% and 95\% credible levels.
    The shaded regions in the 1D PDFs and the quoted parameter values correspond to the 68\% credible intervals.
    }
    \label{fig:corner_cut_patchy_fixed_f}
\end{figure}

For completeness, we show the PDFs and contour maps of the parameters of interest in the baseline PFDM analysis.
These parameters include the FDM mass and the IGM thermal parameters, $T_0$ and $u_0$, at each redshift.
The result is shown in Figure~\ref{fig:corner_cut_patchy_fixed_f}. 
At all three redshifts, there is a clear degeneracy between $m_{\mathrm{FDM}}$ and $u_0$ around
$\log_{10}(m_{\mathrm{FDM}}/\mathrm{eV})\simeq -20.8$, with lower $m_{\mathrm{FDM}}$ corresponding to lower $u_0$.
This indicates that, for $\log_{10}(m_{\mathrm{FDM}}/\mathrm{eV})\simeq -20.8$, the effect of the 1D flux power spectrum caused by PFDM can be partly compensated by weaker pressure smoothing.
For lower FDM masses, the imprint of PFDM on the 1D flux power spectrum becomes too strong to be absorbed by changes in the thermal parameters, and this region of parameter space is therefore disfavoured by our results.

\FloatBarrier

\bibliographystyle{JHEP}
\bibliography{biblio.bib}

\providecommand{\href}[2]{#2}\begingroup\raggedright\begin{thebibliography}{100}

\bibitem{2020A&A...641A...6P}
{Planck Collaboration}, N.~{Aghanim}, Y.~{Akrami}, M.~{Ashdown}, J.~{Aumont},
  C.~{Baccigalupi} et~al., \emph{{Planck 2018 results. VI. Cosmological
  parameters}}, \href{https://doi.org/10.1051/0004-6361/201833910}{\emph{\aap}
  {\bfseries 641} (2020) A6}
  [\href{https://arxiv.org/abs/1807.06209}{{\ttfamily 1807.06209}}].

\bibitem{2020A&A...641A...1P}
{Planck Collaboration}, N.~{Aghanim}, Y.~{Akrami}, F.~{Arroja}, M.~{Ashdown},
  J.~{Aumont} et~al., \emph{{Planck 2018 results. I. Overview and the
  cosmological legacy of Planck}},
  \href{https://doi.org/10.1051/0004-6361/201833880}{\emph{\aap} {\bfseries
  641} (2020) A1} [\href{https://arxiv.org/abs/1807.06205}{{\ttfamily
  1807.06205}}].

\bibitem{1970ApJ...159..379R}
V.C.~{Rubin} and W.K.~{Ford}, Jr., \emph{{Rotation of the Andromeda Nebula from
  a Spectroscopic Survey of Emission Regions}},
  \href{https://doi.org/10.1086/150317}{\emph{\apj} {\bfseries 159} (1970)
  379}.

\bibitem{1985ApJ...295..305V}
T.S.~{van Albada}, J.N.~{Bahcall}, K.~{Begeman} and R.~{Sancisi},
  \emph{{Distribution of dark matter in the spiral galaxy NGC 3198.}},
  \href{https://doi.org/10.1086/163375}{\emph{\apj} {\bfseries 295} (1985)
  305}.

\bibitem{2015PASJ...67...75S}
Y.~{Sofue}, \emph{{Dark halos of M 31 and the Milky Way}},
  \href{https://doi.org/10.1093/pasj/psv042}{\emph{\pasj} {\bfseries 67} (2015)
  75} [\href{https://arxiv.org/abs/1504.05368}{{\ttfamily 1504.05368}}].

\bibitem{1996PhR...267..195J}
G.~{Jungman}, M.~{Kamionkowski} and K.~{Griest}, \emph{{Supersymmetric dark
  matter}}, \href{https://doi.org/10.1016/0370-1573(95)00058-5}{\emph{\physrep}
  {\bfseries 267} (1996) 195}
  [\href{https://arxiv.org/abs/hep-ph/9506380}{{\ttfamily hep-ph/9506380}}].

\bibitem{2005Natur.435..629S}
V.~{Springel}, S.D.M.~{White}, A.~{Jenkins}, C.S.~{Frenk}, N.~{Yoshida},
  L.~{Gao} et~al., \emph{{Simulations of the formation, evolution and
  clustering of galaxies and quasars}},
  \href{https://doi.org/10.1038/nature03597}{\emph{\nat} {\bfseries 435} (2005)
  629} [\href{https://arxiv.org/abs/astro-ph/0504097}{{\ttfamily
  astro-ph/0504097}}].

\bibitem{2012AnP...524..507F}
C.S.~{Frenk} and S.D.M.~{White}, \emph{{Dark matter and cosmic structure}},
  \href{https://doi.org/10.1002/andp.201200212}{\emph{Annalen der Physik}
  {\bfseries 524} (2012) 507}
  [\href{https://arxiv.org/abs/1210.0544}{{\ttfamily 1210.0544}}].

\bibitem{2014MNRAS.444.1518V}
M.~{Vogelsberger}, S.~{Genel}, V.~{Springel}, P.~{Torrey}, D.~{Sijacki},
  D.~{Xu} et~al., \emph{{Introducing the Illustris Project: simulating the
  coevolution of dark and visible matter in the Universe}},
  \href{https://doi.org/10.1093/mnras/stu1536}{\emph{\mnras} {\bfseries 444}
  (2014) 1518} [\href{https://arxiv.org/abs/1405.2921}{{\ttfamily 1405.2921}}].

\bibitem{2018MNRAS.475..624N}
D.~{Nelson}, A.~{Pillepich}, V.~{Springel}, R.~{Weinberger}, L.~{Hernquist},
  R.~{Pakmor} et~al., \emph{{First results from the IllustrisTNG simulations:
  the galaxy colour bimodality}},
  \href{https://doi.org/10.1093/mnras/stx3040}{\emph{\mnras} {\bfseries 475}
  (2018) 624} [\href{https://arxiv.org/abs/1707.03395}{{\ttfamily
  1707.03395}}].

\bibitem{2025JCAP...11..080T}
T.~{Totani}, \emph{{20 GeV halo-like excess of the Galactic diffuse emission
  and implications for dark matter annihilation}},
  \href{https://doi.org/10.1088/1475-7516/2025/11/080}{\emph{\jcap} {\bfseries
  2025} (2025) 080} [\href{https://arxiv.org/abs/2507.07209}{{\ttfamily
  2507.07209}}].

\bibitem{2025arXiv251212176W}
X.~{Wang} and K.-K.~{Duan}, \emph{{Constraining the dark matter origin of the
  halo-like 20 GeV $\gamma$-ray excess with the AMS-02 antiproton data}},
  \href{https://doi.org/10.48550/arXiv.2512.12176}{\emph{arXiv e-prints} (2025)
  arXiv:2512.12176} [\href{https://arxiv.org/abs/2512.12176}{{\ttfamily
  2512.12176}}].

\bibitem{2017Galax...5...17D}
A.~{Del Popolo} and M.~{Le Delliou}, \emph{{Small Scale Problems of the
  {\ensuremath{\Lambda}}CDM Model: A Short Review}},
  \href{https://doi.org/10.3390/galaxies5010017}{\emph{Galaxies} {\bfseries 5}
  (2017) 17} [\href{https://arxiv.org/abs/1606.07790}{{\ttfamily 1606.07790}}].

\bibitem{2017ARA&A..55..343B}
J.S.~{Bullock} and M.~{Boylan-Kolchin}, \emph{{Small-Scale Challenges to the
  {\ensuremath{\Lambda}}CDM Paradigm}},
  \href{https://doi.org/10.1146/annurev-astro-091916-055313}{\emph{\araa}
  {\bfseries 55} (2017) 343}
  [\href{https://arxiv.org/abs/1707.04256}{{\ttfamily 1707.04256}}].

\bibitem{2026NatAs..10..440V}
S.~{Vegetti}, S.D.M.~{White}, J.P.~{McKean}, D.M.~{Powell}, C.~{Spingola},
  D.~{Massari} et~al., \emph{{A possible challenge for cold and warm dark
  matter}}, \href{https://doi.org/10.1038/s41550-025-02746-w}{\emph{Nature
  Astronomy} {\bfseries 10} (2026) 440}.

\bibitem{2026arXiv260116818H}
S.~{Hou}, S.~{Xiang}, Y.-L.~{Sming Tsai}, D.~{Yang}, Y.~{Shu}, N.~{Li} et~al.,
  \emph{{Flux-ratio anomalies in cusp quasars reveal dark matter beyond CDM}},
  \href{https://doi.org/10.48550/arXiv.2601.16818}{\emph{arXiv e-prints} (2026)
  arXiv:2601.16818} [\href{https://arxiv.org/abs/2601.16818}{{\ttfamily
  2601.16818}}].

\bibitem{2000PhRvL..85.1158H}
W.~{Hu}, R.~{Barkana} and A.~{Gruzinov}, \emph{{Fuzzy Cold Dark Matter: The
  Wave Properties of Ultralight Particles}},
  \href{https://doi.org/10.1103/PhysRevLett.85.1158}{\emph{\prl} {\bfseries 85}
  (2000) 1158} [\href{https://arxiv.org/abs/astro-ph/0003365}{{\ttfamily
  astro-ph/0003365}}].

\bibitem{2016PhR...643....1M}
D.J.E.~{Marsh}, \emph{{Axion cosmology}},
  \href{https://doi.org/10.1016/j.physrep.2016.06.005}{\emph{\physrep}
  {\bfseries 643} (2016) 1} [\href{https://arxiv.org/abs/1510.07633}{{\ttfamily
  1510.07633}}].

\bibitem{2021ARA&A..59..247H}
L.~{Hui}, \emph{{Wave Dark Matter}},
  \href{https://doi.org/10.1146/annurev-astro-120920-010024}{\emph{\araa}
  {\bfseries 59} (2021) 247}
  [\href{https://arxiv.org/abs/2101.11735}{{\ttfamily 2101.11735}}].

\bibitem{2025arXiv250700705E}
A.~{Eberhardt} and E.G.M.~{Ferreira}, \emph{{Ultralight fuzzy dark matter
  review}}, \href{https://doi.org/10.48550/arXiv.2507.00705}{\emph{arXiv
  e-prints} (2025) arXiv:2507.00705}
  [\href{https://arxiv.org/abs/2507.00705}{{\ttfamily 2507.00705}}].

\bibitem{2019PrPNP.104....1B}
A.~{Boyarsky}, M.~{Drewes}, T.~{Lasserre}, S.~{Mertens} and O.~{Ruchayskiy},
  \emph{{Sterile neutrino Dark Matter}},
  \href{https://doi.org/10.1016/j.ppnp.2018.07.004}{\emph{Progress in Particle
  and Nuclear Physics} {\bfseries 104} (2019) 1}
  [\href{https://arxiv.org/abs/1807.07938}{{\ttfamily 1807.07938}}].

\bibitem{2001PhLB..518....8B}
C.~{B{\oe}hm}, P.~{Fayet} and R.~{Schaeffer}, \emph{{Constraining dark matter
  candidates from structure formation}},
  \href{https://doi.org/10.1016/S0370-2693(01)01060-7}{\emph{Physics Letters B}
  {\bfseries 518} (2001) 8}
  [\href{https://arxiv.org/abs/astro-ph/0012504}{{\ttfamily
  astro-ph/0012504}}].

\bibitem{2018PhRvL.121h1301G}
V.~{Gluscevic} and K.K.~{Boddy}, \emph{{Constraints on Scattering of keV-TeV
  Dark Matter with Protons in the Early Universe}},
  \href{https://doi.org/10.1103/PhysRevLett.121.081301}{\emph{\prl} {\bfseries
  121} (2018) 081301} [\href{https://arxiv.org/abs/1712.07133}{{\ttfamily
  1712.07133}}].

\bibitem{2018PhR...730....1T}
S.~{Tulin} and H.-B.~{Yu}, \emph{{Dark matter self-interactions and small scale
  structure}},
  \href{https://doi.org/10.1016/j.physrep.2017.11.004}{\emph{\physrep}
  {\bfseries 730} (2018) 1} [\href{https://arxiv.org/abs/1705.02358}{{\ttfamily
  1705.02358}}].

\bibitem{2025RvMP...97d5004A}
S.~{Adhikari}, A.~{Banerjee}, K.K.~{Boddy}, F.-Y.~{Cyr-Racine}, H.~{Desmond},
  C.~{Dvorkin} et~al., \emph{{Astrophysical tests of dark matter
  self-interactions}}, \href{https://doi.org/10.1103/m2vm-59y3}{\emph{Reviews
  of Modern Physics} {\bfseries 97} (2025) 045004}
  [\href{https://arxiv.org/abs/2207.10638}{{\ttfamily 2207.10638}}].

\bibitem{2016ApJ...818...89S}
H.-Y.~{Schive}, T.~{Chiueh}, T.~{Broadhurst} and K.-W.~{Huang},
  \emph{{Contrasting Galaxy Formation from Quantum Wave Dark Matter,
  {\ensuremath{\psi}}DM, with {\ensuremath{\Lambda}}CDM, using Planck and
  Hubble Data}}, \href{https://doi.org/10.3847/0004-637X/818/1/89}{\emph{\apj}
  {\bfseries 818} (2016) 89}
  [\href{https://arxiv.org/abs/1508.04621}{{\ttfamily 1508.04621}}].

\bibitem{2022PhRvD.105l3529P}
S.~{Passaglia} and W.~{Hu}, \emph{{Accurate effective fluid approximation for
  ultralight axions}},
  \href{https://doi.org/10.1103/PhysRevD.105.123529}{\emph{\prd} {\bfseries
  105} (2022) 123529} [\href{https://arxiv.org/abs/2201.10238}{{\ttfamily
  2201.10238}}].

\bibitem{2023MNRAS.524.4256M}
S.~{May} and V.~{Springel}, \emph{{The halo mass function and filaments in full
  cosmological simulations with fuzzy dark matter}},
  \href{https://doi.org/10.1093/mnras/stad2031}{\emph{\mnras} {\bfseries 524}
  (2023) 4256} [\href{https://arxiv.org/abs/2209.14886}{{\ttfamily
  2209.14886}}].

\bibitem{2020MNRAS.492.5721D}
E.Y.~{Davies} and P.~{Mocz}, \emph{{Fuzzy dark matter soliton cores around
  supermassive black holes}},
  \href{https://doi.org/10.1093/mnras/staa202}{\emph{\mnras} {\bfseries 492}
  (2020) 5721} [\href{https://arxiv.org/abs/1908.04790}{{\ttfamily
  1908.04790}}].

\bibitem{2019PhRvL.123e1103M}
D.J.E.~{Marsh} and J.C.~{Niemeyer}, \emph{{Strong Constraints on Fuzzy Dark
  Matter from Ultrafaint Dwarf Galaxy Eridanus II}},
  \href{https://doi.org/10.1103/PhysRevLett.123.051103}{\emph{\prl} {\bfseries
  123} (2019) 051103} [\href{https://arxiv.org/abs/1810.08543}{{\ttfamily
  1810.08543}}].

\bibitem{2021PhRvL.126g1302R}
K.K.~{Rogers} and H.V.~{Peiris}, \emph{{Strong Bound on Canonical Ultralight
  Axion Dark Matter from the Lyman-Alpha Forest}},
  \href{https://doi.org/10.1103/PhysRevLett.126.071302}{\emph{\prl} {\bfseries
  126} (2021) 071302} [\href{https://arxiv.org/abs/2007.12705}{{\ttfamily
  2007.12705}}].

\bibitem{2025PhRvL.134o1001Z}
T.~{Zimmermann}, J.~{Alvey}, D.J.E.~{Marsh}, M.~{Fairbairn} and J.I.~{Read},
  \emph{{Dwarf Galaxies Imply Dark Matter is Heavier than 2.2{\texttimes}10-21
  eV}}, \href{https://doi.org/10.1103/PhysRevLett.134.151001}{\emph{\prl}
  {\bfseries 134} (2025) 151001}
  [\href{https://arxiv.org/abs/2405.20374}{{\ttfamily 2405.20374}}].

\bibitem{2025ApJ...986..127N}
E.O.~{Nadler}, R.~{An}, V.~{Gluscevic}, A.~{Benson} and X.~{Du}, \emph{{COZMIC.
  I. Cosmological Zoom-in Simulations with Initial Conditions Beyond Cold Dark
  Matter}}, \href{https://doi.org/10.3847/1538-4357/adceef}{\emph{\apj}
  {\bfseries 986} (2025) 127}
  [\href{https://arxiv.org/abs/2410.03635}{{\ttfamily 2410.03635}}].

\bibitem{2026ApJ..1000...88L}
J.~{Liu}, Y.~{Gong} and K.~{Liao}, \emph{{Joint Constraints on Fuzzy and Warm
  Dark Matter from Satellite Populations of the Milky Way and Andromeda}},
  \href{https://doi.org/10.3847/1538-4357/ae48e9}{\emph{\apj} {\bfseries 1000}
  (2026) 88} [\href{https://arxiv.org/abs/2512.01361}{{\ttfamily 2512.01361}}].

\bibitem{2022PhRvD.106f3517D}
N.~{Dalal} and A.~{Kravtsov}, \emph{{Excluding fuzzy dark matter with sizes and
  stellar kinematics of ultrafaint dwarf galaxies}},
  \href{https://doi.org/10.1103/PhysRevD.106.063517}{\emph{\prd} {\bfseries
  106} (2022) 063517} [\href{https://arxiv.org/abs/2203.05750}{{\ttfamily
  2203.05750}}].

\bibitem{2023MNRAS.521.2608M}
P.~{Mocz}, A.~{Fialkov}, M.~{Vogelsberger}, M.~{Boylan-Kolchin},
  P.-H.~{Chavanis}, M.A.~{Amin} et~al., \emph{{Cosmological structure formation
  and soliton phase transition in fuzzy dark matter with axion
  self-interactions}},
  \href{https://doi.org/10.1093/mnras/stad694}{\emph{\mnras} {\bfseries 521}
  (2023) 2608} [\href{https://arxiv.org/abs/2301.10266}{{\ttfamily
  2301.10266}}].

\bibitem{2025arXiv250902781M}
S.~{May}, N.~{Dalal} and A.~{Kravtsov}, \emph{{Updated bounds on ultra-light
  dark matter from the tiniest galaxies}},
  \href{https://doi.org/10.48550/arXiv.2509.02781}{\emph{arXiv e-prints} (2025)
  arXiv:2509.02781} [\href{https://arxiv.org/abs/2509.02781}{{\ttfamily
  2509.02781}}].

\bibitem{2025MNRAS.540.2653C}
H.Y.J.~{Chan}, H.-Y.~{Schive}, V.H.~{Robles}, A.~{Kunkel}, G.-M.~{Su} and
  P.-Y.~{Liao}, \emph{{Cosmological zoom-in simulation of fuzzy dark matter
  down to z = 0: tidal evolution of subhaloes in a Milky Way-sized halo}},
  \href{https://doi.org/10.1093/mnras/staf828}{\emph{\mnras} {\bfseries 540}
  (2025) 2653} [\href{https://arxiv.org/abs/2504.10387}{{\ttfamily
  2504.10387}}].

\bibitem{2026arXiv260307175W}
D.~{Wardana}, K.~{Hayashi}, M.~{Chiba} and E.G.M.~{Ferreira}, \emph{{Fuzzy Dark
  Matter and the Impact of Core-Halo Diversity on Its Particle Mass
  Constraints}}, \href{https://doi.org/10.48550/arXiv.2603.07175}{\emph{arXiv
  e-prints} (2026) arXiv:2603.07175}
  [\href{https://arxiv.org/abs/2603.07175}{{\ttfamily 2603.07175}}].

\bibitem{2026arXiv260426393L}
Y.~{Liu} and X.~{Li}, \emph{{Tidal Heating of Stellar Clusters in Fuzzy Dark
  Matter Halos}}, \href{https://doi.org/10.48550/arXiv.2604.26393}{\emph{arXiv
  e-prints} (2026) arXiv:2604.26393}
  [\href{https://arxiv.org/abs/2604.26393}{{\ttfamily 2604.26393}}].

\bibitem{2026PhRvD.113b3055T}
L.~{Teodori}, A.~{Caputo} and K.~{Blum}, \emph{{Ultralight dark matter
  simulations and stellar dynamics: Tension in dwarf galaxies for
  m5{\texttimes}10-21 eV}},
  \href{https://doi.org/10.1103/jc6p-rlvh}{\emph{\prd} {\bfseries 113} (2026)
  023055} [\href{https://arxiv.org/abs/2501.07631}{{\ttfamily 2501.07631}}].

\bibitem{2026arXiv260401278C}
A.~{Caputo} and L.~{Teodori}, \emph{{Influence of tides and self-gravity on
  Ultra-Light Dark Matter Bounds from Dwarf Galaxies}},
  \href{https://doi.org/10.48550/arXiv.2604.01278}{\emph{arXiv e-prints} (2026)
  arXiv:2604.01278} [\href{https://arxiv.org/abs/2604.01278}{{\ttfamily
  2604.01278}}].

\bibitem{2023PhRvD.107h3014G}
M.~{Gosenca}, A.~{Eberhardt}, Y.~{Wang}, B.~{Eggemeier}, E.~{Kendall},
  J.L.~{Zagorac} et~al., \emph{{Multifield ultralight dark matter}},
  \href{https://doi.org/10.1103/PhysRevD.107.083014}{\emph{\prd} {\bfseries
  107} (2023) 083014} [\href{https://arxiv.org/abs/2301.07114}{{\ttfamily
  2301.07114}}].

\bibitem{2023MNRAS.524L..84P}
D.M.~{Powell}, S.~{Vegetti}, J.P.~{McKean}, S.D.M.~{White}, E.G.M.~{Ferreira},
  S.~{May} et~al., \emph{{A lensed radio jet at milli-arcsecond resolution -
  II. Constraints on fuzzy dark matter from an extended gravitational arc}},
  \href{https://doi.org/10.1093/mnrasl/slad074}{\emph{\mnras} {\bfseries 524}
  (2023) L84} [\href{https://arxiv.org/abs/2302.10941}{{\ttfamily
  2302.10941}}].

\bibitem{2023JCAP...06..023R}
K.K.~{Rogers}, R.~{Hlo{\v{z}}ek}, A.~{Lagu{\"e}}, M.M.~{Ivanov},
  O.H.E.~{Philcox}, G.~{Cabass} et~al., \emph{{Ultra-light axions and the S
  $_{8}$ tension: joint constraints from the cosmic microwave background and
  galaxy clustering}},
  \href{https://doi.org/10.1088/1475-7516/2023/06/023}{\emph{\jcap} {\bfseries
  2023} (2023) 023} [\href{https://arxiv.org/abs/2301.08361}{{\ttfamily
  2301.08361}}].

\bibitem{2024ApJ...976...40W}
H.~{Winch}, K.K.~{Rogers}, R.~{Hlo{\v{z}}ek} and D.J.E.~{Marsh},
  \emph{{High-redshift, Small-scale Tests of Ultralight Axion Dark Matter Using
  Hubble and Webb Galaxy UV Luminosities}},
  \href{https://doi.org/10.3847/1538-4357/ad7a73}{\emph{\apj} {\bfseries 976}
  (2024) 40} [\href{https://arxiv.org/abs/2404.11071}{{\ttfamily 2404.11071}}].

\bibitem{2024PhRvD.110l3532L}
H.~{Lazare}, J.~{Flitter} and E.D.~{Kovetz}, \emph{{Constraints on the fuzzy
  dark matter mass window from high-redshift observables}},
  \href{https://doi.org/10.1103/PhysRevD.110.123532}{\emph{\prd} {\bfseries
  110} (2024) 123532} [\href{https://arxiv.org/abs/2407.19549}{{\ttfamily
  2407.19549}}].

\bibitem{2026arXiv260610006C}
W.~{Crumrine}, D.~{Pang}, E.O.~{Nadler}, A.~{Benson} and V.~{Gluscevic},
  \emph{{Mixed Dark Matter: Limits from the Milky Way Satellite Galaxies}},
  \href{https://doi.org/10.48550/arXiv.2606.10006}{\emph{arXiv e-prints} (2026)
  arXiv:2606.10006} [\href{https://arxiv.org/abs/2606.10006}{{\ttfamily
  2606.10006}}].

\bibitem{2026arXiv260606410L}
A.~{Lagu{\"e}}, K.K.~{Rogers}, M.S.~{Madhavacheril}, J.R.~{Bond},
  E.~{Calabrese}, M.J.~{Devlin} et~al., \emph{{The Atacama Cosmology Telescope:
  Probing new signatures of ultralight axions with gravitational lensing}},
  \href{https://doi.org/10.48550/arXiv.2606.06410}{\emph{arXiv e-prints} (2026)
  arXiv:2606.06410} [\href{https://arxiv.org/abs/2606.06410}{{\ttfamily
  2606.06410}}].

\bibitem{2026JCAP...01..047V}
F.~{Verdiani}, E.~{Castorina}, E.~{Salvioni} and E.~{Sefusatti}, \emph{{The
  Effective Field Theory of Large Scale Structure for mixed dark matter
  scenarios}},
  \href{https://doi.org/10.1088/1475-7516/2026/01/047}{\emph{\jcap} {\bfseries
  2026} (2026) 047} [\href{https://arxiv.org/abs/2507.08792}{{\ttfamily
  2507.08792}}].

\bibitem{2016ARA&A..54..313M}
M.~{McQuinn}, \emph{{The Evolution of the Intergalactic Medium}},
  \href{https://doi.org/10.1146/annurev-astro-082214-122355}{\emph{\araa}
  {\bfseries 54} (2016) 313}
  [\href{https://arxiv.org/abs/1512.00086}{{\ttfamily 1512.00086}}].

\bibitem{2026enap....4..401T}
N.~{Tejos}, \emph{{The intergalactic medium}},  in \emph{Encyclopedia of
  Astrophysics, Volume 4}, vol.~4, pp.~401--432, Jan., 2026,
  \href{https://doi.org/10.1016/B978-0-443-21439-4.00087-0}{DOI}
  [\href{https://arxiv.org/abs/2504.12539}{{\ttfamily 2504.12539}}].

\bibitem{2013PhRvD..88d3502V}
M.~{Viel}, G.D.~{Becker}, J.S.~{Bolton} and M.G.~{Haehnelt}, \emph{{Warm dark
  matter as a solution to the small scale crisis: New constraints from high
  redshift Lyman-$\alpha$ forest data}},
  \href{https://doi.org/10.1103/PhysRevD.88.043502}{\emph{\prd} {\bfseries 88}
  (2013) 043502} [\href{https://arxiv.org/abs/1306.2314}{{\ttfamily
  1306.2314}}].

\bibitem{2017PhLB..773..258G}
A.~{Garzilli}, A.~{Boyarsky} and O.~{Ruchayskiy}, \emph{{Cutoff in the
  Lyman-{\ensuremath{\alpha}} forest power spectrum: Warm IGM or warm dark
  matter?}},
  \href{https://doi.org/10.1016/j.physletb.2017.08.022}{\emph{Physics Letters
  B} {\bfseries 773} (2017) 258}
  [\href{https://arxiv.org/abs/1510.07006}{{\ttfamily 1510.07006}}].

\bibitem{2017PhRvD..96b3522I}
V.~{Ir{\v{s}}i{\v{c}}}, M.~{Viel}, M.G.~{Haehnelt}, J.S.~{Bolton},
  S.~{Cristiani}, G.D.~{Becker} et~al., \emph{{New constraints on the
  free-streaming of warm dark matter from intermediate and small scale
  Lyman-$\alpha$ forest data}},
  \href{https://doi.org/10.1103/PhysRevD.96.023522}{\emph{\prd} {\bfseries 96}
  (2017) 023522} [\href{https://arxiv.org/abs/1702.01764}{{\ttfamily
  1702.01764}}].

\bibitem{2017MNRAS.471.4606A}
E.~{Armengaud}, N.~{Palanque-Delabrouille}, C.~{Y{\`e}che}, D.J.E.~{Marsh} and
  J.~{Baur}, \emph{{Constraining the mass of light bosonic dark matter using
  SDSS Lyman-$\alpha$ forest}},
  \href{https://doi.org/10.1093/mnras/stx1870}{\emph{\mnras} {\bfseries 471}
  (2017) 4606} [\href{https://arxiv.org/abs/1703.09126}{{\ttfamily
  1703.09126}}].

\bibitem{2017PhRvD..96l3514K}
T.~{Kobayashi}, R.~{Murgia}, A.~{De Simone}, V.~{Ir{\v{s}}i{\v{c}}} and
  M.~{Viel}, \emph{{Lyman-$\alpha$ constraints on ultralight scalar dark
  matter: Implications for the early and late universe}},
  \href{https://doi.org/10.1103/PhysRevD.96.123514}{\emph{\prd} {\bfseries 96}
  (2017) 123514} [\href{https://arxiv.org/abs/1708.00015}{{\ttfamily
  1708.00015}}].

\bibitem{2017PhRvL.119c1302I}
V.~{Ir{\v{s}}i{\v{c}}}, M.~{Viel}, M.G.~{Haehnelt}, J.S.~{Bolton} and
  G.D.~{Becker}, \emph{{First Constraints on Fuzzy Dark Matter from
  Lyman-$\alpha$ Forest Data and Hydrodynamical Simulations}},
  \href{https://doi.org/10.1103/PhysRevLett.119.031302}{\emph{\prl} {\bfseries
  119} (2017) 031302} [\href{https://arxiv.org/abs/1703.04683}{{\ttfamily
  1703.04683}}].

\bibitem{2018PhRvD..98h3540M}
R.~{Murgia}, V.~{Ir{\v{s}}i{\v{c}}} and M.~{Viel}, \emph{{Novel constraints on
  noncold, nonthermal dark matter from Lyman-$\alpha$ forest data}},
  \href{https://doi.org/10.1103/PhysRevD.98.083540}{\emph{\prd} {\bfseries 98}
  (2018) 083540} [\href{https://arxiv.org/abs/1806.08371}{{\ttfamily
  1806.08371}}].

\bibitem{2022JCAP...10..032H}
D.C.~{Hooper}, N.~{Sch{\"o}neberg}, R.~{Murgia}, M.~{Archidiacono},
  J.~{Lesgourgues} and M.~{Viel}, \emph{{One likelihood to bind them all:
  Lyman-$\alpha$ constraints on non-standard dark matter}},
  \href{https://doi.org/10.1088/1475-7516/2022/10/032}{\emph{\jcap} {\bfseries
  2022} (2022) 032} [\href{https://arxiv.org/abs/2206.08188}{{\ttfamily
  2206.08188}}].

\bibitem{2023PhRvD.108b3502V}
B.~{Villasenor}, B.~{Robertson}, P.~{Madau} and E.~{Schneider}, \emph{{New
  constraints on warm dark matter from the Lyman-$\alpha$ forest power
  spectrum}}, \href{https://doi.org/10.1103/PhysRevD.108.023502}{\emph{\prd}
  {\bfseries 108} (2023) 023502}
  [\href{https://arxiv.org/abs/2209.14220}{{\ttfamily 2209.14220}}].

\bibitem{2024PhRvD.109d3511I}
V.~{Ir{\v{s}}i{\v{c}}}, M.~{Viel}, M.G.~{Haehnelt}, J.S.~{Bolton}, M.~{Molaro},
  E.~{Puchwein} et~al., \emph{{Unveiling dark matter free streaming at the
  smallest scales with the high redshift Lyman-alpha forest}},
  \href{https://doi.org/10.1103/PhysRevD.109.043511}{\emph{\prd} {\bfseries
  109} (2024) 043511} [\href{https://arxiv.org/abs/2309.04533}{{\ttfamily
  2309.04533}}].

\bibitem{2025PhRvD.112d3502G}
O.~{Garcia-Gallego}, V.~{Ir{\v{s}}i{\v{c}}}, M.G.~{Haehnelt}, M.~{Viel} and
  J.S.~{Bolton}, \emph{{Constraining mixed dark matter models with
  high-redshift Lyman-alpha forest data}},
  \href{https://doi.org/10.1103/4k29-h99l}{\emph{\prd} {\bfseries 112} (2025)
  043502} [\href{https://arxiv.org/abs/2504.06367}{{\ttfamily 2504.06367}}].

\bibitem{2026arXiv260324331Z}
S.-Y.~{Zhao}, Y.-C.~{Dai}, W.~{Liao} and Y.-S.~{Lu}, \emph{{Lyman-$\alpha$
  Forest Constraint on Dark Matter from Dark Sector Decay}},
  \href{https://doi.org/10.48550/arXiv.2603.24331}{\emph{arXiv e-prints} (2026)
  arXiv:2603.24331} [\href{https://arxiv.org/abs/2603.24331}{{\ttfamily
  2603.24331}}].

\bibitem{2026arXiv260304401G}
O.~{Garcia-Gallego}, V.~{Ir{\v{s}}i{\v{c}}}, M.~{Viel}, M.G.~{Haehnelt} and
  J.S.~{Bolton}, \emph{{Post-inflationary axion constraints from the
  Lyman-$\alpha$ forest}},
  \href{https://doi.org/10.48550/arXiv.2603.04401}{\emph{arXiv e-prints} (2026)
  arXiv:2603.04401} [\href{https://arxiv.org/abs/2603.04401}{{\ttfamily
  2603.04401}}].

\bibitem{2024PhRvD.109b3507I}
M.M.~{Ivanov}, \emph{{Lyman alpha forest power spectrum in effective field
  theory}}, \href{https://doi.org/10.1103/PhysRevD.109.023507}{\emph{\prd}
  {\bfseries 109} (2024) 023507}
  [\href{https://arxiv.org/abs/2309.10133}{{\ttfamily 2309.10133}}].

\bibitem{2026PhRvL.136q1402I}
M.M.~{Ivanov} and S.~{Trifinopoulos}, \emph{{Effective Field Theory Constraints
  on Primordial Black Holes from the High-Redshift Lyman-{\ensuremath{\alpha}}
  Forest}}, \href{https://doi.org/10.1103/8g8z-bmxd}{\emph{\prl} {\bfseries
  136} (2026) 171402} [\href{https://arxiv.org/abs/2508.04767}{{\ttfamily
  2508.04767}}].

\bibitem{2019ApJ...872..101B}
E.~{Boera}, G.D.~{Becker}, J.S.~{Bolton} and F.~{Nasir}, \emph{{Revealing
  Reionization with the Thermal History of the Intergalactic Medium: New
  Constraints from the Ly$\alpha$ Flux Power Spectrum}},
  \href{https://doi.org/10.3847/1538-4357/aafee4}{\emph{\apj} {\bfseries 872}
  (2019) 101} [\href{https://arxiv.org/abs/1809.06980}{{\ttfamily
  1809.06980}}].

\bibitem{2000SPIE.4008..534D}
H.~{Dekker}, S.~{D'Odorico}, A.~{Kaufer}, B.~{Delabre} and H.~{Kotzlowski},
  \emph{{Design, construction, and performance of UVES, the echelle
  spectrograph for the UT2 Kueyen Telescope at the ESO Paranal Observatory}},
  in \emph{Optical and IR Telescope Instrumentation and Detectors}, M.~{Iye}
  and A.F.~{Moorwood}, eds., vol.~4008 of \emph{Society of Photo-Optical
  Instrumentation Engineers (SPIE) Conference Series}, pp.~534--545, Aug.,
  2000, \href{https://doi.org/10.1117/12.395512}{DOI}.

\bibitem{1994SPIE.2198..362V}
S.S.~{Vogt}, S.L.~{Allen}, B.C.~{Bigelow}, L.~{Bresee}, B.~{Brown},
  T.~{Cantrall} et~al., \emph{{HIRES: the high-resolution echelle spectrometer
  on the Keck 10-m Telescope}},  in \emph{Instrumentation in Astronomy VIII},
  D.L.~{Crawford} and E.R.~{Craine}, eds., vol.~2198 of \emph{Society of
  Photo-Optical Instrumentation Engineers (SPIE) Conference Series}, p.~362,
  June, 1994, \href{https://doi.org/10.1117/12.176725}{DOI}.

\bibitem{1997ApJ...486..599H}
L.~{Hui}, N.Y.~{Gnedin} and Y.~{Zhang}, \emph{{The Statistics of Density Peaks
  and the Column Density Distribution of the Ly{\ensuremath{\alpha}} Forest}},
  \href{https://doi.org/10.1086/304539}{\emph{\apj} {\bfseries 486} (1997) 599}
  [\href{https://arxiv.org/abs/astro-ph/9608157}{{\ttfamily
  astro-ph/9608157}}].

\bibitem{1997MNRAS.292...27H}
L.~{Hui} and N.Y.~{Gnedin}, \emph{{Equation of state of the photoionized
  intergalactic medium}},
  \href{https://doi.org/10.1093/mnras/292.1.27}{\emph{\mnras} {\bfseries 292}
  (1997) 27} [\href{https://arxiv.org/abs/astro-ph/9612232}{{\ttfamily
  astro-ph/9612232}}].

\bibitem{1998MNRAS.296...44G}
N.Y.~{Gnedin} and L.~{Hui}, \emph{{Probing the Universe with the Lyalpha forest
  - I. Hydrodynamics of the low-density intergalactic medium}},
  \href{https://doi.org/10.1046/j.1365-8711.1998.01249.x}{\emph{\mnras}
  {\bfseries 296} (1998) 44}
  [\href{https://arxiv.org/abs/astro-ph/9706219}{{\ttfamily
  astro-ph/9706219}}].

\bibitem{2002MNRAS.334..107G}
N.Y.~{Gnedin} and A.J.S.~{Hamilton}, \emph{{Matter power spectrum from the
  Lyman-alpha forest: myth or reality?}},
  \href{https://doi.org/10.1046/j.1365-8711.2002.05490.x}{\emph{\mnras}
  {\bfseries 334} (2002) 107}
  [\href{https://arxiv.org/abs/astro-ph/0111194}{{\ttfamily
  astro-ph/0111194}}].

\bibitem{2003ApJ...583..525G}
N.Y.~{Gnedin}, E.J.~{Baker}, T.J.~{Bethell}, M.M.~{Drosback}, A.G.~{Harford},
  A.K.~{Hicks} et~al., \emph{{Linear Gas Dynamics in the Expanding Universe}},
  \href{https://doi.org/10.1086/345424}{\emph{\apj} {\bfseries 583} (2003) 525}
  [\href{https://arxiv.org/abs/astro-ph/0206421}{{\ttfamily
  astro-ph/0206421}}].

\bibitem{2015ApJ...812...30K}
G.~{Kulkarni}, J.F.~{Hennawi}, J.~{O{\~n}orbe}, A.~{Rorai} and V.~{Springel},
  \emph{{Characterizing the Pressure Smoothing Scale of the Intergalactic
  Medium}}, \href{https://doi.org/10.1088/0004-637X/812/1/30}{\emph{\apj}
  {\bfseries 812} (2015) 30}
  [\href{https://arxiv.org/abs/1504.00366}{{\ttfamily 1504.00366}}].

\bibitem{2016MNRAS.463.2335N}
F.~{Nasir}, J.S.~{Bolton} and G.D.~{Becker}, \emph{{Inferring the IGM thermal
  history during reionization with the Lyman-$\alpha$ forest power spectrum at
  redshift z $\simeq$ 5}},
  \href{https://doi.org/10.1093/mnras/stw2147}{\emph{\mnras} {\bfseries 463}
  (2016) 2335} [\href{https://arxiv.org/abs/1605.04155}{{\ttfamily
  1605.04155}}].

\bibitem{2007MNRAS.374..493B}
J.S.~{Bolton} and M.G.~{Haehnelt}, \emph{{The nature and evolution of the
  highly ionized near-zones in the absorption spectra of $z \sim 6$ quasars}},
  \href{https://doi.org/10.1111/j.1365-2966.2006.11176.x}{\emph{\mnras}
  {\bfseries 374} (2007) 493}
  [\href{https://arxiv.org/abs/astro-ph/0607331}{{\ttfamily
  astro-ph/0607331}}].

\bibitem{2005MNRAS.357.1178B}
J.S.~{Bolton}, M.G.~{Haehnelt}, M.~{Viel} and V.~{Springel}, \emph{{The Lyman
  {\ensuremath{\alpha}} forest opacity and the metagalactic hydrogen ionization
  rate at $z \sim 2-4$}},
  \href{https://doi.org/10.1111/j.1365-2966.2005.08704.x}{\emph{\mnras}
  {\bfseries 357} (2005) 1178}
  [\href{https://arxiv.org/abs/astro-ph/0411072}{{\ttfamily
  astro-ph/0411072}}].

\bibitem{2001ApJ...556...93B}
P.~{Bode}, J.P.~{Ostriker} and N.~{Turok}, \emph{{Halo Formation in Warm Dark
  Matter Models}}, \href{https://doi.org/10.1086/321541}{\emph{\apj} {\bfseries
  556} (2001) 93} [\href{https://arxiv.org/abs/astro-ph/0010389}{{\ttfamily
  astro-ph/0010389}}].

\bibitem{2019PhRvD..99f3509L}
X.~{Li}, L.~{Hui} and G.L.~{Bryan}, \emph{{Numerical and perturbative
  computations of the fuzzy dark matter model}},
  \href{https://doi.org/10.1103/PhysRevD.99.063509}{\emph{\prd} {\bfseries 99}
  (2019) 063509} [\href{https://arxiv.org/abs/1810.01915}{{\ttfamily
  1810.01915}}].

\bibitem{2019MNRAS.482.3227N}
M.~{Nori}, R.~{Murgia}, V.~{Ir{\v{s}}i{\v{c}}}, M.~{Baldi} and M.~{Viel},
  \emph{{Lyman-$\alpha$ forest and non-linear structure characterization in
  Fuzzy Dark Matter cosmologies}},
  \href{https://doi.org/10.1093/mnras/sty2888}{\emph{\mnras} {\bfseries 482}
  (2019) 3227} [\href{https://arxiv.org/abs/1809.09619}{{\ttfamily
  1809.09619}}].

\bibitem{2026arXiv260406038W}
Y.F.~{Wang}, \emph{{Lyman-$\alpha$ Forest Signatures of Mixed Fuzzy and Cold
  Dark Matter}}, \href{https://doi.org/10.48550/arXiv.2604.06038}{\emph{arXiv
  e-prints} (2026) arXiv:2604.06038}
  [\href{https://arxiv.org/abs/2604.06038}{{\ttfamily 2604.06038}}].

\bibitem{2021PhRvD.103d3526R}
K.K.~{Rogers} and H.V.~{Peiris}, \emph{{General framework for cosmological dark
  matter bounds using N -body simulations}},
  \href{https://doi.org/10.1103/PhysRevD.103.043526}{\emph{\prd} {\bfseries
  103} (2021) 043526} [\href{https://arxiv.org/abs/2007.13751}{{\ttfamily
  2007.13751}}].

\bibitem{yu_feng_2018_1451799}
Y.~Feng, S.~Bird, L.~Anderson, A.~Font-Ribera and C.~Pedersen,
  \emph{Mp-gadget/mp-gadget: A tag for getting a doi},  Oct., 2018.
\newblock 10.5281/zenodo.1451799.

\bibitem{2022MNRAS.512.3703B}
S.~{Bird}, Y.~{Ni}, T.~{Di Matteo}, R.~{Croft}, Y.~{Feng} and N.~{Chen},
  \emph{{The ASTRID simulation: galaxy formation and reionization}},
  \href{https://doi.org/10.1093/mnras/stac648}{\emph{\mnras} {\bfseries 512}
  (2022) 3703} [\href{https://arxiv.org/abs/2111.01160}{{\ttfamily
  2111.01160}}].

\bibitem{2022MNRAS.513..670N}
Y.~{Ni}, T.~{Di Matteo}, S.~{Bird}, R.~{Croft}, Y.~{Feng}, N.~{Chen} et~al.,
  \emph{{The ASTRID simulation: the evolution of supermassive black holes}},
  \href{https://doi.org/10.1093/mnras/stac351}{\emph{\mnras} {\bfseries 513}
  (2022) 670} [\href{https://arxiv.org/abs/2110.14154}{{\ttfamily
  2110.14154}}].

\bibitem{2026arXiv260628482Y}
Z.~{Yuan}, C.~{Gemmell}, K.K.~{Rogers}, J.~{Barron}, S.~{Roy}, D.~{Curtin}
  et~al., \emph{{Strongest constraints on dark acoustic oscillations from the
  Lyman-alpha forest}},
  \href{https://doi.org/10.48550/arXiv.2606.28482}{\emph{arXiv e-prints} (2026)
  arXiv:2606.28482} [\href{https://arxiv.org/abs/2606.28482}{{\ttfamily
  2606.28482}}].

\bibitem{2020JCAP...06..002B}
S.~{Bird}, Y.~{Feng}, C.~{Pedersen} and A.~{Font-Ribera}, \emph{{More accurate
  simulations with separate initial conditions for baryons and dark matter}},
  \href{https://doi.org/10.1088/1475-7516/2020/06/002}{\emph{\jcap} {\bfseries
  2020} (2020) 002} [\href{https://arxiv.org/abs/2002.00015}{{\ttfamily
  2002.00015}}].

\bibitem{2000ApJ...538..473L}
A.~{Lewis}, A.~{Challinor} and A.~{Lasenby}, \emph{{Efficient Computation of
  Cosmic Microwave Background Anisotropies in Closed Friedmann-Robertson-Walker
  Models}}, \href{https://doi.org/10.1086/309179}{\emph{\apj} {\bfseries 538}
  (2000) 473} [\href{https://arxiv.org/abs/astro-ph/9911177}{{\ttfamily
  astro-ph/9911177}}].

\bibitem{2015PhRvD..91j3512H}
R.~{Hlozek}, D.~{Grin}, D.J.E.~{Marsh} and P.G.~{Ferreira}, \emph{{A search for
  ultralight axions using precision cosmological data}},
  \href{https://doi.org/10.1103/PhysRevD.91.103512}{\emph{\prd} {\bfseries 91}
  (2015) 103512} [\href{https://arxiv.org/abs/1410.2896}{{\ttfamily
  1410.2896}}].

\bibitem{2004MNRAS.354..684V}
M.~{Viel}, M.G.~{Haehnelt} and V.~{Springel}, \emph{{Inferring the dark matter
  power spectrum from the Lyman {\ensuremath{\alpha}} forest in high-resolution
  QSO absorption spectra}},
  \href{https://doi.org/10.1111/j.1365-2966.2004.08224.x}{\emph{\mnras}
  {\bfseries 354} (2004) 684}
  [\href{https://arxiv.org/abs/astro-ph/0404600}{{\ttfamily
  astro-ph/0404600}}].

\bibitem{2012ApJ...746..125H}
F.~{Haardt} and P.~{Madau}, \emph{{Radiative Transfer in a Clumpy Universe. IV.
  New Synthesis Models of the Cosmic UV/X-Ray Background}},
  \href{https://doi.org/10.1088/0004-637X/746/2/125}{\emph{\apj} {\bfseries
  746} (2012) 125}.

\bibitem{2017ApJ...837..106O}
J.~{O{\~n}orbe}, J.F.~{Hennawi} and Z.~{Luki{\'c}}, \emph{{Self-consistent
  Modeling of Reionization in Cosmological Hydrodynamical Simulations}},
  \href{https://doi.org/10.3847/1538-4357/aa6031}{\emph{\apj} {\bfseries 837}
  (2017) 106} [\href{https://arxiv.org/abs/1607.04218}{{\ttfamily
  1607.04218}}].

\bibitem{2017ascl.soft10012B}
S.~{Bird}, ``{FSFE: Fake Spectra Flux Extractor}.'' Astrophysics Source Code
  Library, record ascl:1710.012, Oct., 2017.

\bibitem{2022MNRAS.509.6119M}
M.~{Molaro}, V.~{Ir{\v{s}}i{\v{c}}}, J.S.~{Bolton}, L.C.~{Keating},
  E.~{Puchwein}, P.~{Gaikwad} et~al., \emph{{The effect of inhomogeneous
  reionization on the Lyman {\ensuremath{\alpha}} forest power spectrum at
  redshift $z > 4$: implications for thermal parameter recovery}},
  \href{https://doi.org/10.1093/mnras/stab3416}{\emph{\mnras} {\bfseries 509}
  (2022) 6119} [\href{https://arxiv.org/abs/2109.06897}{{\ttfamily
  2109.06897}}].

\bibitem{2026MNRAS.546f2262M}
K.~{Ma}, J.S.~{Bolton}, V.~{Ir{\v{s}}i{\v{c}}}, P.~{Gaikwad} and E.~{Puchwein},
  \emph{{An improved model for the effect of correlated Si III absorption on
  the one-dimensional Lyman-{\ensuremath{\alpha}} forest power spectrum}},
  \href{https://doi.org/10.1093/mnras/staf2262}{\emph{\mnras} {\bfseries 546}
  (2026) staf2262} [\href{https://arxiv.org/abs/2509.08613}{{\ttfamily
  2509.08613}}].

\bibitem{2019MNRAS.490.1588G}
P.~{Gaikwad}, R.~{Srianand}, V.~{Khaire} and T.R.~{Choudhury}, \emph{{Effect of
  non-equilibrium ionization on derived physical conditions of the high-z
  intergalactic medium}},
  \href{https://doi.org/10.1093/mnras/stz2692}{\emph{\mnras} {\bfseries 490}
  (2019) 1588} [\href{https://arxiv.org/abs/1812.01016}{{\ttfamily
  1812.01016}}].

\bibitem{mckay2000comparison}
M.D.~McKay, R.J.~Beckman and W.J.~Conover, \emph{A comparison of three methods
  for selecting values of input variables in the analysis of output from a
  computer code}, {\emph{Technometrics} {\bfseries 42} (2000) 55}.

\bibitem{2018MNRAS.478.3935N}
M.~{Nori} and M.~{Baldi}, \emph{{AX-GADGET: a new code for cosmological
  simulations of Fuzzy Dark Matter and Axion models}},
  \href{https://doi.org/10.1093/mnras/sty1224}{\emph{\mnras} {\bfseries 478}
  (2018) 3935} [\href{https://arxiv.org/abs/1801.08144}{{\ttfamily
  1801.08144}}].

\bibitem{2019JCAP...02..050B}
S.~{Bird}, K.K.~{Rogers}, H.V.~{Peiris}, L.~{Verde}, A.~{Font-Ribera} and
  A.~{Pontzen}, \emph{{An emulator for the Lyman-{\ensuremath{\alpha}}
  forest}}, \href{https://doi.org/10.1088/1475-7516/2019/02/050}{\emph{\jcap}
  {\bfseries 2019} (2019) 050}
  [\href{https://arxiv.org/abs/1812.04654}{{\ttfamily 1812.04654}}].

\bibitem{2019JCAP...02..031R}
K.K.~{Rogers}, H.V.~{Peiris}, A.~{Pontzen}, S.~{Bird}, L.~{Verde} and
  A.~{Font-Ribera}, \emph{{Bayesian emulator optimisation for cosmology:
  application to the Lyman-alpha forest}},
  \href{https://doi.org/10.1088/1475-7516/2019/02/031}{\emph{\jcap} {\bfseries
  2019} (2019) 031} [\href{https://arxiv.org/abs/1812.04631}{{\ttfamily
  1812.04631}}].

\bibitem{2022PhRvL.128q1301R}
K.K.~{Rogers}, C.~{Dvorkin} and H.V.~{Peiris}, \emph{{Limits on the Light Dark
  Matter-Proton Cross Section from Cosmic Large-Scale Structure}},
  \href{https://doi.org/10.1103/PhysRevLett.128.171301}{\emph{\prl} {\bfseries
  128} (2022) 171301} [\href{https://arxiv.org/abs/2111.10386}{{\ttfamily
  2111.10386}}].

\bibitem{2017arXiv170203118E}
S.~{Elfwing}, E.~{Uchibe} and K.~{Doya}, \emph{{Sigmoid-Weighted Linear Units
  for Neural Network Function Approximation in Reinforcement Learning}},
  \href{https://doi.org/10.48550/arXiv.1702.03118}{\emph{arXiv e-prints} (2017)
  arXiv:1702.03118} [\href{https://arxiv.org/abs/1702.03118}{{\ttfamily
  1702.03118}}].

\bibitem{2017arXiv171105101L}
I.~{Loshchilov} and F.~{Hutter}, \emph{{Decoupled Weight Decay
  Regularization}},
  \href{https://doi.org/10.48550/arXiv.1711.05101}{\emph{arXiv e-prints} (2017)
  arXiv:1711.05101} [\href{https://arxiv.org/abs/1711.05101}{{\ttfamily
  1711.05101}}].

\bibitem{2013PASP..125..306F}
D.~{Foreman-Mackey}, D.W.~{Hogg}, D.~{Lang} and J.~{Goodman}, \emph{{emcee: The
  MCMC Hammer}}, \href{https://doi.org/10.1086/670067}{\emph{\pasp} {\bfseries
  125} (2013) 306} [\href{https://arxiv.org/abs/1202.3665}{{\ttfamily
  1202.3665}}].

\bibitem{Buhmann_Dyn_1993}
M.~Buhmann and N.~Dyn, \emph{Spectral convergence of multiquadric
  interpolation},
  \href{https://doi.org/10.1017/S0013091500018411}{\emph{Proceedings of the
  Edinburgh Mathematical Society} {\bfseries 36} (1993) 319–333}.

\bibitem{2024MNRAS.530.4920K}
N.K.~{Khan}, G.~{Kulkarni}, J.S.~{Bolton}, M.G.~{Haehnelt},
  V.~{Ir{\v{s}}i{\v{c}}}, E.~{Puchwein} et~al., \emph{{Particle initialization
  effects on Lyman-{\ensuremath{\alpha}} forest statistics in cosmological SPH
  simulations}}, \href{https://doi.org/10.1093/mnras/stae662}{\emph{\mnras}
  {\bfseries 530} (2024) 4920}
  [\href{https://arxiv.org/abs/2310.07767}{{\ttfamily 2310.07767}}].

\bibitem{2025arXiv251000107G}
O.~{Garcia-Gallego}, V.~{Ir{\v{s}}i{\v{c}}}, M.G.~{Haehnelt} and J.S.~{Bolton},
  \emph{{Constraints on the Thompson optical depth to the CMB from the
  Lyman-$\alpha$ forest}},
  \href{https://doi.org/10.48550/arXiv.2510.00107}{\emph{arXiv e-prints} (2025)
  arXiv:2510.00107} [\href{https://arxiv.org/abs/2510.00107}{{\ttfamily
  2510.00107}}].

\bibitem{2025PhRvD.111h3504H}
L.~{Herold}, E.G.M.~{Ferreira} and L.~{Heinrich}, \emph{{Profile likelihoods in
  cosmology: When, why, and how illustrated with {\ensuremath{\Lambda}}CDM,
  massive neutrinos, and dark energy}},
  \href{https://doi.org/10.1103/PhysRevD.111.083504}{\emph{\prd} {\bfseries
  111} (2025) 083504} [\href{https://arxiv.org/abs/2408.07700}{{\ttfamily
  2408.07700}}].

\bibitem{2026arXiv260325731S}
M.C.~{Straight}, T.~{Karwal}, J.L.~{Bernal} and K.K.~{Boddy}, \emph{{CMB
  constraints on dark matter-proton scattering: investigating prior-volume
  effects using profile likelihoods}},
  \href{https://doi.org/10.48550/arXiv.2603.25731}{\emph{arXiv e-prints} (2026)
  arXiv:2603.25731} [\href{https://arxiv.org/abs/2603.25731}{{\ttfamily
  2603.25731}}].

\bibitem{2021MNRAS.506.5848E}
W.~{Enzi}, R.~{Murgia}, O.~{Newton}, S.~{Vegetti}, C.~{Frenk}, M.~{Viel}
  et~al., \emph{{Joint constraints on thermal relic dark matter from strong
  gravitational lensing, the Ly {\ensuremath{\alpha}} forest, and Milky Way
  satellites}}, \href{https://doi.org/10.1093/mnras/stab1960}{\emph{\mnras}
  {\bfseries 506} (2021) 5848}
  [\href{https://arxiv.org/abs/2010.13802}{{\ttfamily 2010.13802}}].

\bibitem{2025PhRvL.135g1001P}
M.~{Pavi{\v{c}}evi{\'c}}, V.~{Ir{\v{s}}i{\v{c}}}, M.~{Viel}, J.S.~{Bolton},
  M.G.~{Haehnelt}, S.~{Martin-Alvarez} et~al., \emph{{Constraints on Primordial
  Magnetic Fields from the Lyman-{\ensuremath{\alpha}} Forest}},
  \href{https://doi.org/10.1103/77rd-vkpz}{\emph{\prl} {\bfseries 135} (2025)
  071001} [\href{https://arxiv.org/abs/2501.06299}{{\ttfamily 2501.06299}}].

\bibitem{Bishop1994MixtureDN}
C.M.~Bishop, \emph{Mixture density networks},  1994,
  \href{https://api.semanticscholar.org/CorpusID:118227751}{https://api.semanticscholar.org/CorpusID:118227751}.

\end{thebibliography}\endgroup



\providecommand{\href}[2]{#2}\begingroup\raggedright\endgroup
\end{document}